\documentclass[letterpaper, 11pt]{amsart}

\usepackage{style} 

\title{Debordering Closure Results in Determinantal and Pfaffian Ideals}

\newif\ifshowauthor
\showauthortrue

\ifshowauthor
\author{Anakin Dey}
\address{Anakin~Dey\\ Department of Mathematics \\ The Ohio State University \\
\href{mailto:dey.92@buckeyemail.osu.edu}
  {{\ttfamily\upshape dey.92@osu.edu}}}
\author{Zeyu Guo}
\address{Zeyu~Guo\\ Department of Computer Science and Engineering \\ The Ohio State University \\
\href{mailto:zguotcs@gmail.com}
  {{\ttfamily\upshape zguotcs@gmail.com}}. Supported by NSF CAREER award CCF-2440926.}
\fi

\newif\ifmynotes
\mynotestrue

\begin{document}

\begin{abstract}
One important question in algebraic complexity is understanding the complexity of polynomial ideals (Grochow, Bulletin of EATCS 131, 2020).
Andrews and Forbes (STOC 2022) studied the determinantal ideals $I^{\det}_{n,m,r}$ generated by the $r\times r$ minors of $n\times m$ matrices.
Over fields of characteristic zero or of sufficiently large characteristic, they showed that for any nonzero $f \in I^{\det}_{n,m,r}$, the determinant of a $t \times t$ matrix of variables with $t = \Theta(r^{1/3})$ is approximately computed by a constant-depth, polynomial-size $f$-oracle algebraic circuit, in the sense that the determinant lies in the border of such circuits.
An analogous result was also obtained for Pfaffians in the same paper.

In this work, we deborder the result of Andrews and Forbes by showing that when $f$ has polynomial degree, the determinant is in fact exactly computed by a constant-depth, polynomial-size $f$-oracle algebraic circuit.
We further establish an analogous result for Pfaffian ideals.

Our results are established using the isolation lemma, combined with a careful analysis of straightening-law expansions of polynomials in determinantal and Pfaffian ideals.
\end{abstract}

\maketitle
\thispagestyle{empty}

\section{Introduction}\label{sec:intro}

Polynomial ideals naturally arise in many parts of algebraic complexity theory.
Understanding their complexity is an important theme, with connections to polynomial identity testing, algebraic natural proofs, and proof complexity; see~\cite{Grochow20} for a survey that explains such connections.

Roughly speaking, the complexity of an ideal $I$ under a given algebraic model (such as algebraic circuits or algebraic branching programs) is defined as the minimum complexity, in that model, of any nonzero polynomial $f \in I$~\cite{Grochow20}.
Proving lower bounds for ideals is harder than for individual polynomials, since one must establish a bound for every nonzero $f \in I$ rather than just a single polynomial.
This motivates the search for a property of the following form: if an arbitrary nonzero $f \in I$ is efficiently computable in a given model, then some polynomial from a distinguished set $S = \set{g_1, \dots, g_k}$ (often a natural generating set of $I$ or some related set with desirable properties) is also efficiently computable in the given model.
In this way, proving lower bounds for arbitrary nonzero $f \in I$ reduces to proving them for $g_1, \dots, g_k$.
We call such a property a \emph{closure result}, by analogy with closure under factorization in the principal ideal case.
In short, closure results reduce the task of showing hardness for all polynomials in $I$ to proving hardness for a small set of distinguished polynomials.

\subsubsection*{Connection with hitting set generators.} 

Closure results for ideals are useful for constructing hitting set generators (HSGs), which are the algebraic-complexity analogs of pseudorandom generators and can be used to derandomize polynomial identity testing (PIT).
The idea is as follows: for $s < n$, we seek a HSG $G\colon \F^s \to \F^n$ against a family $\mathcal{C}$ of $n$-variate polynomials, meaning that $G$ has the property that for any nonzero polynomial $f \in \mathcal{C}$, we have $f \circ G \neq 0$.
Here, one should think of $\mathcal{C}$ as a family of ``low-complexity'' polynomials.
Let $I$ be the ideal of polynomials $f$ satisfying $f \circ G = 0$, also known as the \emph{vanishing ideal} of $G$~\cite{HvMM24}.
To show that $G$ has the desired HSG property, it suffices to show that every nonzero $f \in I$ has sufficiently high complexity so that no such $f$ can belong to $\mathcal{C}$.
With a closure result for $I$, this reduces to proving lower bounds for the distinguished set of polynomials $\set{g_1, \dots, g_k}$.
This can be viewed as a particular way of implementing the ``hardness vs. randomness'' paradigm of~\cite{NW94} in the setting of algebraic complexity.

\subsubsection*{Principal ideals.}  

The simplest polynomial ideals are the \emph{principal} ideals, \ie, the ideals generated by a single polynomial $g$.  
For such an ideal $I$, any nonzero $f \in I$ is a multiple of $g$ and questions may be rephrased in the context of factorization.
Thus, choosing $S = \set{g}$, the closure result described above follows from the classical closure result under factorization, which states that if $f$ is efficiently computable in a given model, then so are its factors.
For the class $\mathbf{VP}$, this property follows from Kaltofen’s work on polynomial factorization, which in particular showed that if $f \in \F[x_1,\dots,x_n]$ is a polynomial of degree $\poly(n)$ computed by an algebraic circuit of size $\poly(n)$, then its factors also admit $\poly(n)$-size algebraic circuits~\cite{Kal86, Kal87, Kal89}.  
Closure under factorization has also been studied, and in some cases established (or partially established), in more restricted models; see~\cite{Oli16,CKS19,BSV20,ST21,DSS22,BKRRSS25}.

Before moving to non-principal ideals, we recall some technical aspects of closure under factorization.
Kaltofen’s result~\cite{Kal86, Kal87, Kal89} more generally states that if $f \in \F[x_1,\dots,x_n]$ is computed by an algebraic circuit of size $s$, then each factor $g$ of $f$ can be computed by an algebraic circuit of size $\poly(n, s, \deg(f))$.  
This polynomial dependence on $\deg(f)$ is not an obstacle when $f$ has degree $\poly(n)$.
However, if $\deg(f)$ is much larger than $\deg(g)$, then the bound $\poly(n, s, \deg(f))$ may be trivial and prevent one from deducing meaningful lower bounds on the complexity of $f$ from those of $g$.  

On the other hand, \Burgisser's \emph{Factor Conjecture}~\cite[Conjecture~8.3]{Bur00} states that over a field of characteristic zero, if $f$ is computed by an algebraic circuit of size $s$, then any factor $g$ of $f$ can be computed by an algebraic circuit of size $\poly(n, s, \deg(g))$ (rather than $\deg(f)$).
Although the conjecture remains open, partial results are known~\cite{Kal86,Bur00,Bur04,DSS22}.
In particular, \Burgisser\ \cite{Bur04} proved that the Factor Conjecture holds when standard circuit complexity is replaced by \emph{border complexity} (also known as \emph{approximative complexity}).
Specifically, \Burgisser\ showed that if $f$ is computed by an algebraic circuit of size $s$, then any factor $g$ of $f$ is \emph{approximately computed} by an algebraic circuit of size polynomial in $n$, $s$, and $\deg(g)$ over the function field $\F(\e)$.
The notion of approximate computation is defined as follows:
\begin{defn}\label{defn:border_complexity}
We say that $g \in \F[x_1,\dots,x_n]$ is \emph{approximately computed b}y a circuit $C$ over $\F(\e)$ if $C$ computes some polynomial $h \in \F(\e)[x_1,\dots,x_n]$, and $h - g \in \e \cdot \F[[\e]][x_1,\dots,x_n]$, where $\F[[\e]]$ denotes the ring of formal power series in $\e$.
\end{defn}

In the settings where $\F[\e][x_1,\dots,x_n]$ can be equipped with the Euclidean topology in a natural way, such as $\F = \C$ or $\R$, this would imply that $h \to g$ as $\e \mapsto 0$.
However, this does not mean that we could just substitute $\e$ by zero in $C$ and obtain a circuit exactly computing $g$, since $C$ may use scalars in $\F(\e)$, such as $1/\e$, that are not well-defined at $\e = 0$.

\subsubsection*{Non-principal ideals.}

Closure results for non-principal ideals were not directly studied until recently.
In~\cite{Grochow20}, Grochow conjectured such a result for determinantal ideals.
Specifically, he conjectured that for any nonzero $f$ in the determinantal ideal $I$ generated by the $\tfrac{n}{2} \times \tfrac{n}{2}$ minors of an $n \times n$ symbolic matrix $X = (x_{i,j})_{1 \leq i,j \leq n}$, the computation of determinants (minors) reduces to the computation of $f$, although the determinant size may be polynomially smaller than $\tfrac{n}{2} \times \tfrac{n}{2}$.
Formally:
\begin{conj}[{\cite[Conjecture~6.3]{Grochow20}}]\label{conj:grochow_det}
Let $X$ be an $n \times n$ matrix of indeterminates and let $I_n$ be the ideal generated by the $\tfrac{n}{2} \times \tfrac{n}{2}$ minors of $X$.
Then for every nonzero polynomial $f \in I_n$, there is a constant-depth\footnote{The original statement of the conjecture does not specify constant-depth algebraic circuits. However, it is already known that the $m \times m$ determinant can be computed by algebraic circuits of size $\poly(m)$.} algebraic circuit of size $\poly(n)$ with $f$-oracle gates that computes the $m \times m$ determinant for some $m = n^{\bigTheta{1}}$.
\end{conj}

\begin{remark}
We note that \Cref{conj:grochow_det} asserts that the computation of $m \times m$ determinants reduces to that of $f$.
The existence of such a reduction is formally stronger than a closure result, and we therefore refer to it as a \emph{reducibility result}.
In this paper we do not carefully distinguish between the two notions, since the main results we prove are also of the stronger reducibility type.
\end{remark}

Analogous to \Burgisser's theorem~\cite{Bur04} that the Factor Conjecture holds with respect to border complexity, Andrews and Forbes~\cite{AF21} proved that Grochow's conjecture holds with respect to border complexity.
Formally, they proved the following theorem:

\begin{thrm}[{\cite[Theorem~1.1]{AF21}}]\label{thrm:AF}
Let $\F$ be a field of characteristic zero.
Let $X = (x_{i,j})_{\substack{1 \leq i \leq n \\ 1 \leq j \leq m}}$ be an $n \times m$ matrix of variables over $\F$ and let $I^{\det}_{n,m,r} \subseteq \F[X] = \F[x_{i,j}]$ be the ideal generated by the $r \times r$ minors of $X$.
Let $f\in  I^{\det}_{n,m,r}$ be a nonzero polynomial.
Then there is a depth-three $f$-oracle circuit of size $\bigO{n^2 m^2}$ that approximately computes the $t \times t$ determinant for $t = \bigTheta{r^{1/3}}$.
\end{thrm}

In the same paper, Andrews and Forbes also gave an analogous result for the Pfaffian of a $2n \times 2n$ skew-symmetric matrix and the ideal generated by Pfaffians of $2r \times 2r$ principal submatrices, and showed that these results continue to hold over fields of sufficiently large characteristic.

As mentioned above, closure results are useful for constructing HSGs and derandomizing PIT by reducing these tasks to proving lower bounds for distinguished polynomials.
In particular, by combining \Cref{thrm:AF} with the celebrated superpolynomial lower bounds for constant-depth algebraic circuits due to Limaye, Srinivasan, and Tavenas~\cite{LST25} (which also hold in the border complexity setting, as observed in~\cite{AF21}), Andrews and Forbes~\cite{AF21} obtained a simple construction of HSGs for constant-depth circuits.
By further composing this construction with scaled copies of itself, they gave a final HSG construction that achieves a near-optimal trade-off between seed length and degree with subpolynomial seed length, improving upon the earlier construction of~\cite{CKS19} and yielding improved subexponential-time deterministic PIT algorithms for constant-depth algebraic circuits.

\subsubsection*{Back to exact computation.}

Note that \Cref{thrm:AF} proves Grochow's conjecture (\Cref{conj:grochow_det}), but only with respect to border complexity.
This leaves open the natural question: Does the conjecture also hold in the standard (exact) setting?

% TODO: Elaborate more on this: does removing dependence on deg(f) imply any strong factor conjecture results?
We begin with a technical issue.
In Grochow's formulation of the conjecture, the algebraic circuit computing the determinant is required to have size $\poly(n)$, independent of $\deg(f)$, even though $\deg(f)$ may be much larger than $n$.
This degree-independent condition is the central technical challenge in proving the conjecture, because existing approaches naturally introduce a dependence on the degree.
The situation is analogous to the difficulty encountered in Bürgisser’s Factor Conjecture, although neither conjecture implies the other, and they could well have different resolutions or require unrelated methods.
%\footnote{In fact, \Cref{conj:grochow_det} appears even more elusive to us than the Factor Conjecture since it is unclear whether Hensel lifting, a technique that is crucial for partial results toward the Factor Conjecture, offers any help in proving \Cref{conj:grochow_det}.} 

On the other hand, most applications in algebraic complexity involve only polynomials of polynomially bounded degree.
For example, Kaltofen’s result showing that $\mathbf{VP}$ is closed under factorization suffices for many purposes, even though the full Factor Conjecture remains open.
Motivated by this, we restrict our attention to polynomials $f$ of degree $\poly(n)$ and ask whether Grochow’s conjecture holds in this regime.
More generally, we may allow circuit size to depend polynomially on both $n$ and $\deg(f)$.
This leads to the following question:
\begin{question}\label{question:main}
Is \Cref{conj:grochow_det} true when $\deg(f)=\poly(n)$? 
More generally, under the notation of \Cref{conj:grochow_det}, is it true that there is a constant-depth algebraic circuit of size $\poly(n,\deg(f))$ (rather than $\poly(n)$ as in \Cref{conj:grochow_det}) with $f$-oracle gates that computes the $t \times t$ determinant for some $t = n^{\bigTheta{1}}$?
\end{question}

\subsection{Main Results}

In this paper, we answer \Cref{question:main} affirmatively by establishing the following theorem:

\begin{thrm}[Informal version of \Cref{thrm:ABP_det_oracle} and \Cref{cor:det_oracle_for_det}]\label{thrm:main}
Let $X = (x_{i,j})_{\substack{1 \leq i \leq n \\ 1 \leq j \leq m}}$ be an $n \times m$ matrix of variables over a field $\F$.
Let $I^{\det}_{n,m,r} \subseteq \F[X] = \F[x_{i,j} \mid 1 \leq i \leq n,\ 1 \leq j \leq m]$ be the ideal generated by the $r \times r$ minors of $X$.  
Let $f \in I^{\det}_{n,m,r}$ be a nonzero polynomial.  
Assume that the characteristic of $\F$ is either zero or greater than $\deg(f)$.
Also assume that the size of $\F$ is sufficiently large.\footnote{Specifically, we require $\abs{\F}$ to be bounded below by some polynomial in $n$, $m$, and $\deg(f)$.
Such assumptions are standard in the literature (for instance, in work on polynomial factorization), since one can often achieve them by passing to a suitable field extension.} 
Then there exists a depth-three $f$-oracle circuit of size $\poly(n, m, \deg(f))$ that computes the $t \times t$ determinant for $t = \bigTheta{r^{1/3}}$.
\end{thrm}

\begin{thrm}[Informal version of \Cref{thrm:ABP_pfaff_oracle} and \Cref{cor:pfaff_oracle_for_pfaff}]\label{thrm:pfaff_main}
Let $X = (x_{i,j})_{1 \leq i < j \leq 2n}$ be a $2n \times 2n$ skew-symmetric matrix of variables over a field $\F$ so that $x_{j, i} = -x_{i, j}$ for $1 \leq j < i \leq n$ and $x_{i, i} = 0$ for $1 \leq i \leq 2n$.
Let $I^{\pfaff}_{2n, 2r} \subseteq \F[X] = \F[x_{i,j} \mid 1 \leq i < j \leq 2n]$ be the ideal generated by the $2r \times 2r$ principal minors of $X$.  
Let $f \in I^{\pfaff}_{2n, 2r}$ be a nonzero polynomial.  
Assume that the characteristic of $\F$ is either zero or greater than $\deg(f)$.
Also assume that the size of $\F$ is sufficiently large. 
Then there exists a depth-three $f$-oracle circuit of size $\poly(n, \deg(f))$ that computes the $t \times t$ Pfaffian of a skew-symmetric matrix for $t = \bigTheta{r^{1/3}}$.
\end{thrm}

It is worth noting that determinantal and Pfaffian ideals can be studied within unified frameworks, notably Hodge algebras~\cite{DEP82} and Standard Monomial Theory~\cite{LR08,Ses16}.
In this paper, however, we do not adopt these approaches as they are more abstract than necessary for our purposes.

While \Cref{thrm:main} does not by itself yield new PIT algorithms, because \Cref{thrm:AF} already suffices given that the lower bounds of~\cite{LST25} hold in the border complexity setting, it resolves a natural theoretical question that was previously open. 
% AD: Fixed this last sentence, see Section 7 of AF22, in particular Theorem 7.3. 
% AD: Do we want to state this here or just save it for a full journal version? I do not know the best way to go about this. I'll leave you to make this decision.

\Cref{thrm:main} may also be viewed as a ``debordering'' result: showing that a statement known to hold with respect to border complexity continues to hold in the exact setting, possibly with somewhat larger complexity bounds.
In general, such results are highly nontrivial: naively converting a circuit that approximately computes a polynomial into one that computes it exactly can cause an exponential blow-up, due to the exponentially large degree in $\e$ of the scalars used in the approximation~\cite{LL89,Bur00}.
Nevertheless, by exploiting the structure of determinantal and Pfaffian ideals, we show that this difficulty can be overcome, yielding debordering results.
See~\cite{DDS22,BDS24,DGIJL24,Shp25} for other nontrivial debordering results in various settings.

\subsection{Proof Overview}

The main tool in our proof is the \emph{isolation lemma}, first introduced by Valiant and Vazirani~\cite{VV86}.
This lemma, along with its derandomized variants, has found numerous applications in theoretical computer science, including parallel algorithms for perfect matching~\cite{MVV87,FGT19,ST17}, polynomial identity testing~\cite{KS01,AMS10,GT20}, and search-to-decision reductions~\cite{VV86,BCGL92,GGR24}, among others.
To the best of our knowledge, however, this paper is the first to apply the isolation lemma to obtain debordering results.

The version of the isolation lemma we use~\cite{KS01} can be stated as follows.
Consider a collection $\mathcal{C}$ of monomials of individual degree at most $K$ in variables $x_1,\dots,x_\ell$.
Suppose we substitute each variable by $x_i \mapsto w^{z_i}$, where the exponents $z_1,\dots,z_\ell$ are chosen independently and uniformly from $\set{0, \dots ,M}$ with $M \geq K\ell/\e$.
Then, with probability at least $1-\e$, there is a unique monomial $\mathfrak{m} \in \mathcal{C}$ that attains the minimum degree in $w$ under this substitution.

In this way, we reduce the number of variables to a single one while ensuring that the polynomial does not vanish after substitution: the unique monomial of minimum degree in $w$ cannot be canceled by others.
For comparison, a Kronecker substitution $x_i \mapsto w^{D^{i-1}}$ for sufficiently large $D$ achieves the same effect.
However, the advantage of the isolation lemma is that the exponents $z_i$ need only be polynomial in $\ell$ and $K$, whereas the Kronecker map requires that $D$ is exponentially large.

To see why this advantage helps us prove \Cref{thrm:main}, let us first review the proof of \Cref{thrm:AF} from~\cite{AF21}.
Let $f \in I^{\det}_{n,m,r}$ be a nonzero polynomial.
Our goal is to compute the $t \times t$ determinant using an $f$-oracle circuit.
The proof in~\cite{AF21} relies on a technique from algebraic combinatorics, specifically \emph{standard monomial theory}, called the straightening law, which expresses $f$ as a linear combination of special polynomials known as standard bideterminants.
Each standard bideterminant is a product of determinants of submatrices of $X = (x_{i,j})$, and in our setting, every such bideterminant includes at least one determinant of sufficiently large size.

The argument then proceeds in two steps.
The first step transforms this linear combination of standard bideterminants into a single ``canonical'' standard bideterminant.
The second step computes a determinant from a standard bideterminant.
Our second step is identical to that in~\cite{AF21}; therefore, we focus on the first step.

At a high level, the way~\cite{AF21} transforms a linear combination of standard bideterminants into a single standard bideterminant is by applying a sequence of linear transformations of the variables, so that each standard bideterminant in the combination is eventually mapped to a term that equals (up to sign) the target standard bideterminant.
However, applying this idea naively does not work, since terms arising from different standard bideterminants may cancel each other.
To avoid this,~\cite{AF21} choose the linear transformations so that they effectively multiply the terms by distinct monomials in some new variables.
As a result, terms originating from different standard bideterminants are ``tagged'' with different monomials.
Finally, a Kronecker-type substitution is applied to merge these new variables into a single one while still maintaining the existence of nonzero terms, yielding a polynomial of the form 
\[
    g = \e^k g_0 + \bigO{\e^{k+1}}
\]
over $\F(\e)$, where $g_0$ is a distinguished standard bideterminant. 
This standard bideterminant can then be used to compute the $s \times s$ determinant that we aim to compute.
The presence of the $\bigO{\e^{k+1}}$ term poses no issue for proving \Cref{thrm:AF}, as the theorem concerns only approximate computation.

To turn approximate computation into exact computation, a natural idea is to extract the degree-$k$ homogeneous component of $g$ in $\e$ and then set $\e = 1$.
However, this approach does not work directly because the degree of $g$ in $\e$ can become exponentially large due to the use of the Kronecker-type substitution.
Consequently, the circuit obtained by extracting the degree-$k$ component directly would have exponential size.

This is precisely where the isolation lemma becomes useful.
By replacing the Kronecker-type substitution with the randomized variable substitution from the isolation lemma, we can ensure that the degree in $\e$ remains only polynomially large.
Consequently, extracting the degree-$k$ homogeneous component yields a polynomial-size circuit, as desired.

Although the high-level idea is simple, carrying it out requires substantial technical work.
Unlike the Kronecker substitution, the isolation lemma offers only limited control over which monomial is isolated.
In particular, the linear transformations from~\cite{AF21} produce additional ``garbage'' terms which, in their setting, were always dominated in the lexicographic order and could be safely ignored.
With the isolation lemma, however, these extraneous terms may instead be isolated—something we must avoid.
Thus, simply combining the isolation lemma with the analysis of~\cite{AF21} does not suffice.
To address this, we integrate the isolation lemma with additional ideas in a subtle way, allowing us to recover the desired terms while discarding the extraneous ones.
The technical details are deferred to \Cref{sec:det}.

\section{Preliminaries}\label{sec:prelims}

For a natural number $n$, we denote the set $\set{1, 2, \ldots, n - 1, n}$ by $[n]$.
We will denote fields by $\F$.
Throughout, we will work primarily with polynomial rings with coefficients in a field.
For a field $\F$ and a set of indeterminates $\overline{x} = \set{x_1, \ldots, x_n}$, we write $\F[\overline{x}]$ for the ring of polynomials in the variables $x_1, \ldots, x_n$ with coefficients in $\F$.
We note that in addition to $\F[\overline{x}]$ being a ring, it is also an $\F$-vector space and that ideals of $\F[\overline{x}]$ are subspaces.
Thus, it will make sense to talk about an $\F$-basis of an ideal in addition to a generating set for that ideal.
Occasionally, we will just mention $\F[\overline{x}]$ without specifying the number of variables when the exact number is not important or is implied by context.
If the variables come from an $n \times m$ matrix $X = (x_{i, j})_{\substack{1 \leq i \leq n \\ 1 \leq j \leq m}}$, then we notate the respective polynomial ring as $\F[X]$.
In any case, when $f \in \F[\overline{x}]$, we write $\deg(f)$ to denote the total degree of $f$.
We will also write $\deg_{x_{i}}(f)$ to denote the total degree of $f$ with respect to $x_{i}$ specifically.

We will assume basic notions in algebraic complexity theory; see, \eg, the survey of Saptharishi~\cite{Sap21}.
In particular, the size of an algebraic circuit is the number of gates plus the number of wires, and scalar multiplications along wires are free.
As we will be working with constant-depth circuits, gates are assumed to have unbounded fan-in and fan-out.

In addition to standard algebraic circuits, we will use algebraic oracle circuits and algebraic branching programs, defined below.
\begin{defn}[Oracle circuit]\label{defn:oracle_ckt}
Let $g(\overline{x}) \in \F[\overline{x}]$, where $\overline{x}$ denotes $x_1, \ldots, x_n$.  
A (algebraic) \emph{$g$-oracle circuit} $C$ over $\F$ is an algebraic circuit augmented with an additional type of gate called a $g$-gate.  
A $g$-gate with inputs $f_1,\dots,f_n$ outputs the polynomial $g(f_1,\dots,f_n)$.
\end{defn}

In our definition of algebraic branching programs, edges are assumed to be labeled by polynomials of degree at most one.
\begin{defn}[Algebraic branching program]\label{defn:ABP}
    An \emph{algebraic branching program} (\emph{ABP}) over $\F[\overline{x}]$ is a directed acyclic graph $A$ such that:
    \begin{itemize}
        \item The nodes $V$ of $A$ are partitioned into non-empty subsets $V = V_0 \sqcup \cdots \sqcup V_d$ for some integer $d\geq 0$ and we call $V_i$ the \emph{$i$-th layer}.
              Layer $V_0$ has exactly one node called the \emph{source $s$} and layer $V_d$ has exactly on node called the \emph{sink $t$}.
        \item Each edge in $A$ goes from layer $V_{i-1}$ to $V_i$ for some $i\in [d]$.
        \item Each edge $e$ in $A$ is labeled with a polynomial $\gamma_e \in \F[\overline{x}]$ of degree at most one.
    \end{itemize}
    For a path $\overline{e} = (e_1, \ldots, e_d)$ from $s$ to $t$, where each edge $e_i$ goes from layer $i - 1$ to layer $i$, define the polynomial $\gamma_{\overline{e}} \defeq \prod_{i = 1}^d \gamma_{e_i}$.
    Then $A$ computes the polynomial $\sum_{\text{path } \overline{e} \text{ from } s \text{ to } t} \gamma_{\overline{e}}$.
\end{defn}
One motivation for the connection to algebraic branching programs is that algebraic branching programs on $n$ nodes compute polynomials which are determinants of $\bigO{n} \times \bigO{n}$ matrices~\cite[Theorem 1]{Val79}.

\subsection{Linear Algebra}\label{sec:linalg}

Throughout, all rings will be commutative with unity.
For any ring $A$ and any natural numbers $n, m$, let $A^{n\times m}$ be the set of all $n\times m$ matrices with entries in $A$.
For such a matrix $M \in A^{n \times m}$, we will write $M = \pqty{m_{i, j}}_{\substack{1 \leq i \leq n \\ 1 \leq j \leq m}}$ to denote that $(i, j)$-th entry of $M$ by $m_{i, j}$ where $1 \leq i \leq n$ and $1 \leq j \leq m$.
For a square matrix $M \in A^{n \times n}$, we will denote the $n \times n$ determinant by $\det_n(M)$.

\begin{defn}[Elementary matrix]\label{defn:elem_mat}
    Let $i,j$ be distinct elements in $[n]$, and let $z \in A$.
    Then the \emph{elementary matrix} $E_{i, j}(z) \in A^{n \times n}$ is an $n \times n$ matrix with $1$'s down the main diagonal, $z$ in position $(i, j)$, and $0$'s elsewhere.
    Note that $\det_n(E_{i, j}(z)) = 1$.

    These elementary matrices correspond to row operations when multiplying on the left, and column operations when multiplying on the right.
    Indeed if $M \in A^{n \times m}$, consider $E_{i, j}(z) \in A^{n \times n}$.
    Then the product $E_{i, j}(z) M$ is the matrix given by adding $z$ times row $j$ of $M$ to row $i$ of $M$.
    Similarly, if we take $E_{i, j}(z) \in A^{m \times m}$, then the product $M E_{i, j}(z)^{\top}$ is the matrix given by adding $z$ times column $j$ of $M$ to column $i$ of $M$.
\end{defn}

Let $A$ be any ring and $M \in A^{n\times m}$.
For subsets $R \subseteq [n]$ and $C \subseteq [m]$, we will denote by $M_{R, C}$ the submatrix of $M$ with rows given by $R$ and columns given by $C$.
If $R = C$ then such a submatrix is \emph{principal}.

For a positive integer $n$, let $S_n$ be the \emph{permutation group} on $[n]$, \ie, the group of bijections $\sigma\colon [n] \to [n]$.
For a permutation $\sigma \in S_n$, we may associate an $n \times n$ \emph{permutation matrix} $C_\sigma$ such that $(C_\sigma)_{i, j} = 1$ if $\sigma(i) = j$ and $0$ otherwise.
Note that $C_\sigma^{-1} = C_\sigma^{\top}$.
The \emph{sign} of a permutation $\sigma \in S_n$ is defined using the number of inversions ($i < j$ such that $\sigma(i) > \sigma(j)$):
\[
  \sgn(\sigma) = (-1)^{\abs{\set{(i, j) \mid 1 \leq i < j \leq n,\ \sigma(i) > \sigma(j)}}} = \pm 1.
\]
Then we have  that $\det_n(C_\sigma) = \sgn(\sigma) = \pm 1$.
For a $n \times n$ matrix $M \in A^{n \times n}$, we have that $(C_\sigma M C_\sigma^{\top})_{i, j} = M_{\sigma(i), \sigma(j)}$.
In other words, $C_\sigma M C_\sigma^{\top}$ is the matrix such that $(C_\sigma M C_\sigma^{\top})_{i, j} = M_{\sigma(i), \sigma(j)}$.

\subsection{Bitableaux}\label{sec:bitab}

\begin{defn}[Partition, Young diagram]\label{defn:partitions}
    A \emph{partition} is a non-increasing sequence of integers $\sigma = (\sigma_1 \geq \sigma_2 \geq \cdots \geq \sigma_k \geq 0)$ for some $k \geq 1$.
    The \emph{size} of a partition is the sum of the size of its parts and is denoted $\abs{\sigma} = \sum_{i = 0}^k \sigma_i$.    
    To such a partition $\sigma$, we can associate the \emph{transpose} $\widehat{\sigma} = (\widehat{\sigma}_1 \geq \cdots \geq \widehat{\sigma}_\ell)$ for some $\ell \geq 1$ where $\widehat{\sigma}_i \defeq \abs{\set{j \in \N | \sigma_j \geq i}}$.
    A partition $\sigma$ can be given pictorially as a \emph{Young diagram} which formally is the set of points $\set{(i, j) | j \leq \sigma_i} \subseteq \N \times \N$.
    This indexing follows the same indexing as entries of a matrix, so that the first coordinate increases as one traverses downwards and the second coordinate increases as one traverses to the right.
    We refer to each point of a Young diagram as a \emph{cell}.
    The Young diagram of $\widehat{\sigma}$ is the reflection of the Young diagram for $\sigma$ over the diagonal $y = x$.
    Note that $\widehat{\sigma}_1$ is the number of non-empty rows of the Young diagram for $\sigma$.
    For example, if $\sigma = (5, 4, 2, 1)$ then $\widehat{\sigma} = (4, 3, 2, 2, 1)$.
    Their Young diagrams can be seen in \Cref{fig:young_diagram_ex}.

    \begin{figure}[H]
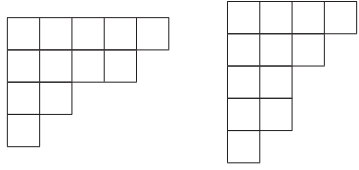

        \centering
        \[
            \Yvcentermath1 \yng(5,4,2,1) \qquad \yng(4,3,2,2,1)
        \]
        \caption{
          The Young diagrams for $\sigma = (5, 4, 2, 1)$ and $\widehat{\sigma} = (4, 3, 2, 2, 1)$.
        }\label{fig:young_diagram_ex}
    \end{figure}
\end{defn}

These partitions have a natural lexicographic ordering which is useful for describing ideals indexed by partitions of a given shape or size.
\begin{defn}\label{defn:lex}
    The ordering $1 < 2 < 3 < \cdots$ induces a lexicographic ordering on partitions, which we denote by $\leqlex$.
    This ordering has a minor caveat in that extending a partition makes it smaller with respect to $\leqlex$.
    Formally, for partitions $\lambda = (\lambda_1 \geq \cdots \geq \lambda_t)$ and $\mu = (\mu_1 \geq \cdots \geq \mu_s)$ we have that $\lambda \leqlex \mu$ if and only if exactly one of the following holds:
    \begin{enumerate}
        \item For some $1 \leq k \leq \min\set{s, t}$, we have that $\lambda_k < \mu_k$ and for all $1 \leq i \leq k - 1$ we have that $\lambda_i = \mu_i$, or
        \item We have that $s \geq t$ and for all $1 \leq i \leq t$ we have that $\lambda_i = \mu_i$, \ie, $\mu$ is a prefix of $\lambda$.
    \end{enumerate}
    \begin{figure}[H]
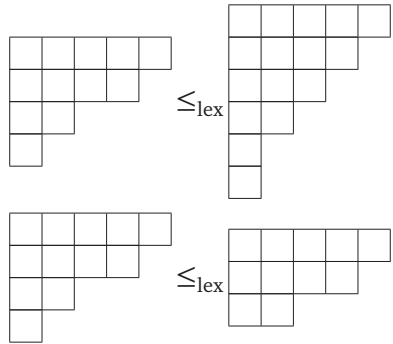

        \centering
        \begin{align*}
            \Yvcentermath1 \yng(5,4,2,1) \leqlex \yng(5,4,3,2,1,1) \\
            \Yvcentermath1 \yng(5,4,2,1) \leqlex \yng(5,4,2)           
        \end{align*}
        \caption{
            Two examples of the order $\leqlex$ on partitions.
            First, we have that $(5, 4, 2, 1) \leqlex (5, 4, 3, 2, 1, 1)$ as $2 < 3$.
            Next, we have that $(5, 4, 2, 1) \leqlex (5, 4, 2)$ as $(5, 4, 2)$ is a prefix of $(5, 4, 2, 1)$.
        }\label{fig:lex_examples}
    \end{figure} 
\end{defn}

\begin{defn}[Young tableau]\label{defn:tableau}
    A \emph{Young tableau}, or \emph{tableau} for short, of shape $\sigma$ is an assignment of positive integers to each cell of the Young diagram of $\sigma$.
    A Young tableau is \emph{semistandard} if its entries are nondecreasing along the rows of the diagram and strictly increasing along the columns.
    A Young tableau is \emph{standard} if its entries are both strictly increasing along the rows of the diagram and strictly increasing along the columns.
    If $T$ is a Young tableau of shape $\sigma$, then its \emph{conjugate} $\widehat{T}$ is a Young tableau of shape $\widehat{\sigma}$ whose entry $T(i, j)$ in cell $(i, j)$ is the entry $T(j, i)$ of cell $(j, i)$ of $T$.
    We will also denote the $i$-th row of $T$ by $T_i$.
    Throughout this paper, we will mostly work with \emph{conjugate semistandard} Young tableaux whose entries are strictly increasing along the rows of the diagram and nondecreasing along the columns, \ie, they are the conjugates of semistandard Young tableaux.
    For example, if $\sigma = (5, 4, 2, 1)$, then a semistandard tableau $T$ for $\sigma$, as well as the conjugate semistandard tableau $\widehat{T}$, are given in \Cref{fig:young_tableau_ex}.
    
    \begin{figure}[H]
        \centering
        \[
            \young(11245,2245,34,5) \qquad \young(1235,124,24,45,5)
        \]
        \caption{
            A Young tableau $T$ for $\sigma = (5, 4, 2, 1)$ as well as its conjugate $\widehat{T}$.
            Note that $T$ is semistandard and $\widehat{T}$ is conjugate semistandard.
        }\label{fig:young_tableau_ex}
    \end{figure}
    We remark that much of the literature on straightening laws refers to what we call a conjugate semistandard Young tableau as just standard or semistandard.
    We stick to our convention as it is the one that is prevalent in more modern papers and texts.
\end{defn}

\begin{defn}\label{defn:size_of_tab}
    Let $\sigma = (\sigma_1 \geq \cdots \geq \sigma_k \geq 0)$ be any partition and $T$ a Young tableau of shape $\sigma$.
    Then we define the \emph{size} of $T$, denoted $\abs{T}$, to be the sum of all entries of $T$:
    \[
        \abs{T} \defeq \sum_{i = 1}^k \sum_{j = 1}^{\sigma_i} T(i, j).
    \]
\end{defn}

In particular, much of our analysis will center around a special set of tableau which for a given shape $\sigma$ are the maximal and minimal conjugate semistandard Young tableau of shape $\sigma$ with respect to the size.
\begin{defn}[Canonical/anti-canonical tableau]\label{defn:canonical_tableau}
    Fix a natural number $n$.
    Let $\sigma = (\sigma_1 \geq \cdots \geq \sigma_k > 0)$ be a partition such that the length $\sigma_i$ of each row is at most $n$.
    Define $K_{\sigma, n}$ to be the tableau of shape $\sigma$ where row $i$ has entries $1, 2, \ldots, \sigma_i$.
    Similarly, define $\overline{K}_{\sigma, n}$ to be the tableau of shape $\sigma$ where row $i$ has entries $(n - \sigma_i + 1, n - \sigma_i + 2, \ldots, n)$.
    We call $K_{\sigma, n}$ and $\overline{K}_{\sigma, n}$ the \emph{canonical} and \emph{anti-canonical} tableau respectively.\footnote{In some sources, such as~\cite{BV88}, these are referred to as the \emph{initial} and \emph{final} tableau.}
    Note that $K_{\sigma, n}$ and $\overline{K}_{\sigma, n}$ are both conjugate semistandard.
    When $n$ is clear from context, we will just write $K_{\sigma}$ and $\overline{K}_{\sigma}$.
    For example, if $\sigma = (5, 4, 2, 1)$ and $n = 7$, then $K_\sigma$ and $\overline{K}_\sigma$ are given in \Cref{fig:canonical_tableau_ex}.
    
    \begin{figure}[H]
        \centering
        \[
            \young(12345,1234,12,1) \qquad \young(34567,4567,67,7)
        \]
        \caption{
          The canonical and anti-canonical tableau for $\sigma = (5, 4, 2, 1)$ and $n = 7$.
        }\label{fig:canonical_tableau_ex}
    \end{figure}
\end{defn}

\subsection{Determinantal Ideals}

Let $\F$ be a field.
Let $X = (x_{i, j})_{\substack{1 \leq i \leq n \\ 1 \leq j \leq m}}$ be a $n \times m$ matrix of variables.
We will denote by $\F[X]$ the polynomial ring $\F[x_{i, j} \mid 1 \leq i \leq n,\ 1 \leq j \leq m]$.
We can natural associate a product of minors of $X$ to a pair of Young tableau where the minors are described by the entries of the Young tableau.
\begin{defn}[Bitableau, bideterminant]\label{defn:bitableau}
    Let $\sigma = (\sigma_1 \geq \cdots \geq \sigma_k)$ be a partition.
    An \emph{$(n, m)$-bitableau} $(S \mid T)$ of shape $\sigma$ is a pair of two Young tableau of shape $\sigma$ with where the entries of $S$ come from $[n]$ and the entries of $T$ come from $[m]$.
    When $n$ and $m$ are clear from context, we just call $(S \mid T)$ a \emph{bitableau}.
    For each $i\in [\widehat{\sigma}_1]$, write the $i$-th row of $S$ and the $i$-th row of $T$ as $S_i = (S(i, 1), \ldots, S(i, \sigma_i))$ and $T_i = (T(i, 1), \ldots,  T(i, \sigma_i))$ respectively.
    The $\sigma_i \times \sigma_i$ minor of $X$ defined by $S_i$ and $T_i$ is the matrix
    \[
        \pqty{x_{S(i, a), T(i, b)}}_{1 \leq a, b \leq \sigma_i} = 
            \begin{pmatrix}
                x_{S(i, 1), T(i, 1)}        & x_{S(i, 1), T(i, 2)}        & \cdots & x_{S(i, 1), T(i, \sigma_i)} \\
                x_{S(i, 2), T(i, 1)}        & x_{S(i, 2), T(i, 2)}        & \cdots & x_{S(i, 2), T(i, \sigma_i)} \\
                \vdots                      & \vdots                      & \ddots & \vdots \\
                x_{S(i, \sigma_i), T(i, 1)} & x_{S(i, \sigma_i), T(i, 2)} & \cdots & x_{S(i, \sigma_i), T(i, \sigma_i)}
            \end{pmatrix}.
    \]
    Then the associated \emph{bideterminant} $(S \mid T)(X)$ is the product of all such minors over $i\in [\widehat{\sigma}_1]$:
    \[
        (S \mid T)(X) \defeq \prod_{i = 1}^{\widehat{\sigma}_1} \det_{\sigma_i}\pqty{x_{S(i, a), T(i, b)}}_{1 \leq a, b \leq \sigma_i}.
    \]
    An $n \times n$ determinant is homogeneous of degree $n$ so that $(S \mid T)(X)$ is a homogeneous polynomial of degree $\abs{\sigma}$.
    For example, if $\sigma = (4, 2, 1)$ and $(S \mid T)$ is the bitableau
    \[
        (S \mid T) = \left(\ \Yvcentermath1 \left. \young(1245,36,4)\ \right\vert\ \young(1356,27,3) \ \right),
    \]
    then the bideterminant $(S \mid T)(X)$ is given by
    \[
        \begin{pmatrix}
            x_{1, 1} & x_{1, 3} & x_{1, 5} & x_{1, 6} \\
            x_{2, 1} & x_{2, 3} & x_{2, 5} & x_{2, 6} \\
            x_{4, 1} & x_{4, 3} & x_{4, 5} & x_{4, 6} \\
            x_{5, 1} & x_{5, 3} & x_{5, 5} & x_{5, 6} 
        \end{pmatrix}
        \begin{pmatrix}
            x_{3, 2} & x_{3, 7} \\
            x_{6, 2} & x_{6, 7}
        \end{pmatrix}
        \begin{pmatrix}
            x_{4, 3}
        \end{pmatrix}.
    \]
    
\end{defn}

The following lemma shows that every monomial in $\F[X]$ can be expressed as a bideterminant, and hence the bideterminants span $\F[X]$ as an $\F$-vector space.
The proof is immediate from the definition.
\begin{lem}\label{lem:det_monomial}
    A degree $d$ monomial $\prod_{i = 1}^d x_{r_i, c_i}$ is given by a bideterminant:
    \[
	   \newcommand{\ra}{r_1}
	   \newcommand{\rb}{r_2}
	   \newcommand{\rd}{r_d}
	   \newcommand{\ca}{c_1}
	   \newcommand{\cb}{c_2}
	   \newcommand{\cd}{c_d}
	   \left(\ \Yvcentermath1 \left.\young(\ra,\rb,\svdots,\rd) \ \right\vert\ \young(\ca,\cb,\svdots,\cd) \ \right)(X) = \prod_{i = 1}^d x_{r_i, c_i}.
    \]
    Thus, the bideterminants span $\F[X]$.
\end{lem}

While the bideterminants span $\F[X]$, there is a specific subset of bideterminants, the \emph{standard} bideterminants, which form an $\F$-basis of $\F[X]$.
\begin{defn}[Standard bitableau, standard bideterminant]\label{defn:det_std}
    We call a bitableau $(S \mid T)$ and its associated bideterminant $(S \mid T)(X)$ \emph{standard} if $S$ and $T$ are both conjugate semistandard Young tableaux.
\end{defn}

The next theorem is known in the literature as the \emph{straightening law}.
\begin{thrm}[{\cite[\S 8, Theorem 3]{DRS74},~\cite[Corollary 2.3]{dCEP80}}]\label{thrm:det_straightening}
    Let $(S \mid T)(X)$ be a bideterminant of shape $\sigma$.
    Then $(S \mid T)(X)$ can be uniquely expressed as a linear combination
    \[
        (S \mid T)(X) = \sum_{(A \mid B)} c_{A, B} (A \mid B)(X),
    \]
    where $(A \mid B)(X)$ is a standard bideterminant of shape $\tau \geq_{\lex} \sigma$, and $c_{A,B} \in \Z$ when $\operatorname{char}(\F) = 0$, while $c_{A,B} \in \Z/p\Z$ when $\operatorname{char}(\F) = p > 0$.
\end{thrm}

The fact that bideterminants are homogeneous polynomials, together with \Cref{lem:det_monomial} and \Cref{thrm:det_straightening}, imply the following corollary.
\begin{cor}\label{cor:det-basis}
The standard bideterminants form an $\F$-basis of $\F[X]$.
In particular, since bideterminants are homogeneous with respect to total degree, the degree-$d$ component of $\F[X]$ has as basis the standard bideterminants of shape $\sigma$ such that $\abs{\sigma} = d$.
\end{cor}

Let $I^{\det}_{n, m, r}$ be the ideal of $\F[X]$ generated by the $r \times r$ minors of $X$.
Note that this ideal is a $\F$-subspace of $\F[X]$.
We have the following crucial statement on the $\F$-basis of this ideal.
\begin{prop}[{\cite[Corollary 1.7]{BC03}}]\label{prop:det_shape}
    Polynomials $f \in I^{\det}_{n, m, r}$ are supported by standard bideterminants of shape $\sigma$ such that $\sigma_1 \geq r$; that is, by those whose first row has length at least $r$.
\end{prop}

The notion of the total degree of a polynomial is too coarse for differentiating terms in the expansion of a polynomial $f \in I^{\det}_{n, m, r}$ with respect to the standard basis.
Thus, we describe a grading of the polynomial ring $\F[X]$ that is better suited for this purpose.
\begin{defn}\label{defn:multideg}
    The polynomial ring $\F[X]$ has a natural $(\N^n \oplus \N^m)$-grading.
    Let $\overline{e}_i \in \N^n$ be the standard basis vector with a $1$ in position $i$ and $0$'s in the other positions.
    Define $\overline{e}_i \in \N^m$ the same way.
    Then we assign $x_{i, j}$ to have degree $\overline{e}_i \oplus \overline{e}_j$.
    We call this assignment the \emph{multidegree}.
    We say a polynomial $f \in \F[X]$ is \emph{multihomogeneous} of multidegree $(s_1 \overline{e}_1 + \cdots + s_n \overline{e_n}) \oplus (t_1 \overline{e}_1 + \cdots + t_m \overline{e}_m)$ if every monomial in $f$ has multidegree $(s_1 \overline{e}_1 + \cdots + s_n \overline{e_n}) \oplus (t_1 \overline{e}_1 + \cdots + t_m \overline{e}_m)$.
\end{defn} 

The following lemma gives the multidegree of bideterminants.
In particular, the multidegree of canonical and anti-canonical bideterminants will allow us to extract terms of a particular shape.
\begin{lem}\label{lem:canonical_degs}
    Fix a natural number $n \in \Z_{\geq 0}$.
    Let $\sigma = (\sigma_1 \geq \cdots \geq \sigma_d)$ be a partition such that the length $\sigma_i$ of each row is at most $n$.
    Let $S, T$ be two conjugate semistandard Young tableaux.
    Then a bideterminant $(S \mid T)(X)$ is multihomogeneous of multidegree $(s_1 \overline{e}_1 + \cdots + s_n \overline{e_n}) \oplus (t_1 \overline{e}_1 + \cdots + t_n \overline{e}_n)$, where $s_i$ is equal to the number of $i$'s that appear in $S$ and $t_j$ is equal to the number of $j$'s that appear in $T$.
    In particular, $(K_\sigma \mid K_\sigma)(X)$ is multihomogeneous of multidegree $(\widehat{\sigma}_1 \overline{e}_1 + \cdots + \widehat{\sigma}_n \overline{e}_1) \oplus (\widehat{\sigma}_1 \overline{e}_1 + \cdots + \widehat{\sigma}_n \overline{e}_1)$.
    Moreover, for distinct partitions $\sigma \neq \tau$, the multidegrees of $(K_\sigma \mid K_\sigma)(X)$ and $(K_\tau \mid K_\tau)(X)$ are distinct, and the same holds for the multidegrees of $(\overline{K}_\sigma \mid \overline{K}_\sigma)(X)$ and $(\overline{K}_\tau \mid \overline{K}_\tau)(X)$. 
\end{lem}
\begin{pf}
    The multidegree of a standard bideterminant $(S \mid T)(X)$ is immediate from \Cref{defn:bitableau}.
    To see that for distinct partitions $\sigma \neq \tau$ the multidegrees of $(K_\sigma \mid K_\sigma)(X)$ and $(K_\tau \mid K_\tau)(X)$ are distinct, we make the following observation.
    The number of $1$'s in $K_{\sigma}$ is the coefficient of $\overline{e}_1$ in the multidegree, the number of $2$'s in $K_{\sigma}$ is the coefficient of $\overline{e}_2$, and so on.
    This allows us to get the values $\widehat{\sigma}_1, \ldots, \widehat{\sigma}_n$ which in turn completely determines $\sigma$.
\end{pf}

\subsection{Pfaffian Ideals} 

We now introduce notation for Pfaffian ideals.
Readers interested only in the classical determinantal case may skip this part.
Let $x_{i, j}$ be indeterminates for $1 \leq i < j \leq 2n$ and let $x_{j, i} \defeq -x_{i, j}$ for $1 \leq j < i \leq 2n$ and $x_{i, i} = 0$ for $1 \leq i \leq 2n$.
Thus, $X = (x_{i, j})_{1 \leq i, j \leq 2n}$ is a skew-symmetric $2n \times 2n$ matrix of variables.
For notational consistency with the prior subsections, we denote the polynomial ring $\F[x_{i, j} \mid 1 \leq i < j \leq 2n]$ by $\F[X]$.

Recall that for a skew-symmetric matrix $X$, the determinant when viewed as a polynomial in the entries of the skew-symmetric matrix is a perfect square of some other polynomial.
The square root of $\det_{2n}(X)$ is called the \emph{Pfaffian} of $X$.
For the Pfaffian of a $2n \times 2n$ matrix $M \in \F^{2n \times 2n}$, we use the notation $\pfaff_{2n}$.
The Pfaffian has a more intrinsic definition given as follows:
\begin{equation}\label{eq:pfaff_defn}
    \pfaff_{2n}(X) = \frac{1}{2^n n!} \sum_{\sigma \in \mathfrak{S}_{2n}} \sgn(\sigma) \prod_{i = 1}^n x_{\sigma(2i - 1), \sigma(2i)}.
\end{equation}
Just like the determinant, this is a polynomial with coefficients $\pm 1$ so that it is well defined over any field.
To see this, note that every monomial in \Cref{eq:pfaff_defn} appears exactly $2^n n!$ times.
We also have that if $m$ is odd, then $\det_m(X) = 0$ when $X$ is a $m \times m$ skew-symmetric matrix so that $\pfaff_m(X) = 0$.
Thus, we only consider Pfaffians on matrices of even order.

Pfaffians are deeply connected to determinants, as indicated by the following lemma:
\begin{lem}[{\cite[Lemma 2.27]{AF21}}]\label{lem:pfaff_conj}
    Let $X$ be a $2n \times 2n$ skew-symmetric matrix and let $E$ be an arbitrary $2n \times 2n$ matrix.
    Then $EXE^\top$ is also skew-symmetric and $\pfaff_{2n}(EXE^\top) = \det_{2n}(E) \pfaff_{2n}(X)$.
\end{lem}

Since the Pfaffian is only well-defined for skew-symmetric matrices, we cannot consider all possible row and column combinations when looking at Pfaffians defined on minors of a matrix.
\begin{defn}[Principal bitableau, bipfaffian]\label{defn:principal_bitableau}
    A bitableau $(S \mid T)$ is \emph{principal} if $S = T$, \ie, if the submatrix of $X$ with rows given by $S$ and columns given by $T$ is principal.
    In this case, we denote $[T] \defeq (T \mid T)$.
    Since $m \times m$ determinants of skew-symmetric matrices are zero for odd $m$, we will assume all partitions have only even row lengths in this case.
    In particular, $\abs{\sigma}$ will always be divisible by $2$.
    
    Let $\sigma$ be a partition such that the length of each row of $\sigma$ is even, and let $T$ be a Young tableau of shape $\sigma$.
    Writing the rows as $T_i = (T(i, 1), \ldots, T(i, \sigma_i))$, the submatrix of $X$ with rows and columns given by $T_i$ is skew-symmetric as well and thus has a well-defined \emph{principal pfaffian}.
    The associated \emph{bipfaffian}\footnote{\cite{AF21} call this a \emph{standard monomial}.
    We deviate from this as standard monomial is a more general term which also refers to standard bideterminants and other analogues in standard monomial theory.} $[T](X)$ is the product of all such Pfaffians defined by the rows of $T$: 
    \[
        [T](X) \defeq \prod_{i = 1}^{\widehat{\sigma}_1} \pfaff\pqty{x_{T(i, a), T(i, b)}}_{1 \leq a, b \leq \sigma_i}.
    \]
    Note that the total degree $\deg([T](X))$ is equal to $\abs{\sigma} / 2$.
\end{defn}

The following lemma shows that every monomial in $\F[X]$ can be expressed as a bipfaffian and hence the bipfaffians span $\F[X]$.
The proof is immediate from the definition.

\begin{lem}\label{lem:pfaff_monomial}
    A degree $d$ monomial $\prod_{i = 1}^d x_{r_i, c_i}$ where for each $1 \leq i \leq d$, $r_{i} < c_{i}$ is given by a bipfaffian:
    \[
	    \newcommand{\ra}{r_1}
	    \newcommand{\rb}{r_2}
	    \newcommand{\rd}{r_d}
	    \newcommand{\ca}{c_1}
	    \newcommand{\cb}{c_2}
	    \newcommand{\cd}{c_d}
	    \left[\ \Yvcentermath1 \young(\ra\ca,\rb\cb,\svdots\svdots,\rd\cd)\ \right](X) = \prod_{i = 1}^d x_{r_i, c_i}.
    \]
\end{lem}

Analogous to the classical determinantal case, a specific subset of bipfaffians form an $\F$-basis of $\F[X]$.
\begin{defn}[Standard bipfaffian]\label{defn:pfaff_std}
    A bipfaffian $[S](X)$ is \emph{standard} if $S$ is a conjugate semistandard Young tableau.
\end{defn}

As in the determinantal case, the Pfaffian case also admits a straightening law.

\begin{thrm}[{\cite[Theorem 6.5]{dCP76},~\cite[\S 5]{HT92}}]\label{thrm:pfaff_straightening}
    Let $S$ be a Young tableau of shape $\sigma$.
    Then $[S](X)$ can be uniquely expressed as a linear combination
    \[
        [S](X) = \sum_{A} c_{A} [A](X),
    \]
    where $[A](X)$ is a standard bipfaffian of shape $\tau \geq_{\lex} \sigma$, and $c_{A} \in \Z$ when $\operatorname{char}(\F) = 0$, while $c_{A} \in \Z/p\Z$ when $\operatorname{char}(\F) = p > 0$.
\end{thrm}

\Cref{lem:pfaff_monomial} and \Cref{thrm:pfaff_straightening} together imply the following corollary.

\begin{cor}\label{cor:pfaff-basis}
The standard bipfaffians form an $\F$-basis of $\F[X]$.
In particular, since bipfaffians are homogeneous with respect to total degree, the degree-$d$ component of $\F[X]$ has as basis the standard bipfaffians of shape $\sigma$ such that $\abs{\sigma} = 2d$.
Note that if $\abs{\sigma} = 2d$, then the number of rows $\widehat{\sigma}_1$ of $\sigma$ is at most $d$ as each non-empty row of $\sigma$ has length at least $2$.
\end{cor}

Let $I^{\pfaff}_{2n, 2r}$ be the ideal of $\F[X]$ generated by the $2r \times 2r$ principal bipfaffians of $X$.
We have the following crucial statement.
\begin{prop}[{\cite[\S 5]{HT92}}]\label{prop:pfaff_shape}
    Polynomials $f \in I^{\pfaff}_{2n, 2r}$ are supported by standard bipfaffians of shape $\sigma$ such that $\sigma_1 \geq 2r$.
\end{prop}

Next, we describe a grading of the polynomial ring $\F[X]$ that is useful for the proof.

\begin{defn}\label{defn:multideg_pfaff}
    The polynomial ring $\F[X]$ has a natural $\N^{2n}$-grading.
    Let $\overline{e}_i \in \N^{2n}$ be the standard basis vector with a $1$ in position $i$ and $0$'s in the other positions.
    Then we assign $x_{i, j}$ to have degree $\overline{e}_i + \overline{e}_j$.
    We call this assignment the \emph{multidegree}.
    We say a polynomial $f \in \F[X]$ is \emph{multihomogeneous} of multidegree $s_1 \overline{e}_1 + \cdots + s_{2n} \overline{e_{2n}}$ if every monomial in $f$ has multidegree $s_1 \overline{e}_1 + \cdots + s_{2n} \overline{e_{2n}}$.
\end{defn} 

The following lemma gives the multidegree of bipfaffians.
It follows immediately from the definition and is proven similarly to \Cref{lem:canonical_degs} and thus we omit the proof.
\begin{lem}\label{lem:canonical_degs_bifpaffian}
    Fix a natural number $n \in \Z_{\geq 0}$.
    Let $\sigma = (\sigma_1 \geq \cdots \geq \sigma_k)$ be a partition such that the length of each row is at most $2n$.
    Let $S$ be a conjugate semistandard Young tableau.
    Then a bipfaffian $[S](X)$ is multihomogeneous of multidegree $s_1 \overline{e}_1 + \cdots + s_{2n} \overline{e}_{2n}$, where $s_i$ is equal to the number of $i$'s that appear in $S$.
    In particular, $[K_\sigma](X)$ is multihomogeneous of multidegree $\widehat{\sigma}_1 \overline{e}_1 + \cdots + \widehat{\sigma}_{2n} \overline{e}_{2n}$.
    Moreover, for distinct partitions $\sigma \neq \tau$, the multidegrees of $[K_\sigma](X)$ and $[K_\tau](X)$ are distinct, and the same holds for the multidegrees of $[\overline{K}_\sigma](X)$ and $[\overline{K}_\tau](X)$.
\end{lem}

\subsection{Isolation Lemma}

One crucial tool used in this work is the \emph{isolation lemma}.
We use the following version due to Klivans and Spielman~\cite{KS01}.

\begin{lem}[{\cite[Lemma 4]{KS01}}]\label{lem:isolation_forms}
    Let $C$ be any collection of distinct linear forms in variables $z_1, \ldots, z_\ell$ with coefficients in the range $\set{0, \ldots, K}$ for some integer $K \in \Z_{\geq 0}$.
    Let $\e > 0$.
    Let $z_1, \ldots, z_\ell$ be independently and uniformly chosen from $\set{0, \ldots, M}$ at random, where $M\geq K \ell / \e$.
    Then, with probability at least $1 - \e$, there is a unique form in $C$ of minimum value at $z_1, \ldots, z_\ell$.
\end{lem}

A simple rephrasing of \Cref{lem:isolation_forms} allows us to isolate a term of a multivariate polynomial $f(y_1,\dots,y_\ell)$ under a random monomial substitution $y_i\mapsto w^{z_i}$:
\begin{cor}\label{cor:isolation_exponents}
    Let $K \in \Z_{\geq 0}$.  
    Let $A$ be any ring, and let $f(y_1, \ldots, y_\ell) \in A[\overline{y}]$ be a polynomial whose individual degree in each variable is at most $K$, \ie, $\deg_{y_i}(f) \leq K$ for all $i \in [\ell]$.  
    Fix $\e > 0$.  
    Choose $z_1, \ldots, z_\ell$ independently and uniformly at random from $\set{0, \ldots, M}$, where $M \geq K\ell/\e$.  
    Let $w$ be a new indeterminate, and define $\phi \colon A[y_1, \ldots, y_\ell] \to A[w]$ to be the ring homomorphism sending $y_i \mapsto w^{z_i}$ for each $i \in [\ell]$.  
    Then, with probability at least $1 - \e$, there exists a unique monomial $\mathfrak{m} = y_1^{e_1}\cdots y_\ell^{e_\ell}$ among all monomials of $f$ such that  $\phi(\mathfrak{m}) = w^{e_1 z_1 + \cdots + e_\ell z_\ell}$
    attains the minimum degree in $w$.
\end{cor}

\subsection{Computing the Homogeneous Components}

There are multiple methods in the literature for computing the homogeneous components of a given polynomial.
In this paper, we use an interpolation-based method and observe that it also applies in the oracle-circuit setting.

\begin{defn}\label{defn:coeff}
For $f\in \F[\overline{x}, t]$ and an integer $i\in\Z_{\geq 0}$, denote by $\coeff_{t^i}(f)$ the coefficient of $t^i$ in $f$ when $f$ is viewed as a univariate polynomial in $t$ over the ring $\F[\overline{x}]$.
Note that $\coeff_{t^i}(f)\in \F[\overline{x}]$.
\end{defn}
The following lemma shows that the coefficients of a polynomial computed by a $g$-oracle circuit $C$ can be extracted by another $g$-oracle circuit of comparable depth. 
While this differs from computing homogeneous components, the two notions are closely related: the degree-$i$ homogeneous component of $f \in \F[x_1,\dots,x_n]$ is exactly $\coeff_{t^i}(f(tx_1,\dots,tx_n))$.
\begin{lem}\label{lem:homogenization}
Let $g \in \F[\overline{x},t]$, and let $f \in \F[\overline{x},t]$ be a degree-$d$ polynomial such that for each $\alpha \in \F$, the polynomial $f(\overline{x},\alpha)$ is computed by a $g$-oracle circuit $C_\alpha$ of size at most $s$ and depth at most $\Delta$.
Assume that $\abs{\F} \geq d + 1$.
Then, for every $0 \leq i \leq d$, there exists a $g$-oracle circuit $D_i$ of size $\bigO{ds}$ and depth $\Delta+1$ that computes $\coeff_{t^i}(f)$, with its top gate being an addition gate and bottom $\Delta$ layers consisting of $d+1$ copies of $g$-oracle circuits of the form $C_\alpha$.
Moreover, if the top gate of each $C_\alpha$ is an addition gate, then the depth bound can be improved to $\Delta$.
\end{lem}
\begin{pf}
    By definition,
    \[
        f = \sum_{i = 0}^d \coeff_{t^i}(f) t^i.
    \]
    Let $\alpha_1, \ldots, \alpha_{d + 1}$ be $d + 1$ distinct elements in $\F$.
    Then we have that 
    \[
        \begin{pmatrix}
            \alpha_1^d & \alpha_1^{d - 1} & \cdots & \alpha_1 & 1 \\ 
            \alpha_2^d & \alpha_2^{d - 1} & \cdots & \alpha_2 & 1 \\ 
            \vdots & \vdots & \ddots & \vdots & \vdots \\
            \alpha_{d}^d & \alpha_{d}^{d - 1} & \cdots & \alpha_{d} & 1 \\ 
            \alpha_{d + 1}^d & \alpha_{d + 1}^{d - 1} & \cdots & \alpha_{d + 1} & 1 
        \end{pmatrix}
        \begin{pmatrix}
            \coeff_{t^d}(f) \\
            \coeff_{t^{d-1}}(f) \\
            \vdots \\
            \coeff_{t}(f) \\
            \coeff_{1}(f)
        \end{pmatrix}
        =
        \begin{pmatrix}
            f(\overline{x},\alpha_1) \\
            f(\overline{x},\alpha_2) \\
            \vdots \\
            f(\overline{x},\alpha_d) \\
            f(\overline{x},\alpha_{d+1})
        \end{pmatrix}        
    \]
    The $(d + 1) \times (d + 1)$ matrix $(\alpha_i^{d + 1 - j}) \in \F^{(d + 1) \times (d + 1)}$ is a Vandermonde matrix.
    As all the $\alpha_i$ are distinct, this matrix is invertible and so there exists a $(d + 1) \times (d + 1)$ matrix $(z_{i, j}) \in \F^{(d + 1) \times (d + 1)}$ such that
    \[
        \begin{pmatrix}
            \coeff_{t^d}(f) \\
            \coeff_{t^{d-1}}(f) \\
            \vdots \\
            \coeff_{t}(f) \\
            \coeff_{1}(f)
        \end{pmatrix}
        =
        \begin{pmatrix}
            z_{1, 1} & z_{1, 2} & \cdots & z_{1, d} & z_{1, d + 1} \\
            z_{2, 1} & z_{2, 2} & \cdots & z_{2, d} & z_{2, d + 1} \\
            \vdots & \vdots & \ddots & \vdots & \vdots \\
            z_{d, 1} & z_{d, 2} & \cdots & z_{d, d} & z_{d, d + 1} \\
            z_{d + 1, 1} & z_{d + 1, 2} & \cdots & z_{d + 1, d} & z_{d + 1, d + 1} \\
        \end{pmatrix}
        \begin{pmatrix}
            f(\overline{x},\alpha_1) \\
            f(\overline{x},\alpha_2) \\
            \vdots \\
            f(\overline{x},\alpha_d) \\
            f(\overline{x},\alpha_{d+1})
        \end{pmatrix} 
    \]
    Fix $0 \leq i \leq d$.
    We now have an explicit expression for $\coeff_{t^i}(f)$:
    \begin{equation}\label{eq:H_i_explicit}
        \coeff_{t^i}(f) = \sum_{j = 1}^{d + 1} z_{d + 1 - i, j} f(\overline{x},\alpha_j).
    \end{equation}
With \Cref{eq:H_i_explicit}, we can now describe the circuit for $\coeff_{t^i}(f)$.
The scalars $\alpha_1, \ldots, \alpha_{d+1}$ are fixed \apriori, and hence so are the $z_{i,j}$ for $i, j \in [d + 1]$.
Recall that scalar multiplications along wires are free.
For each $j \in [d + 1]$, computing $f(\overline{x},\alpha_j)$ requires a single $g$-oracle circuit $C_{\alpha_j}$ of size at most $s$ and depth $\Delta$.
Hence, evaluating all $d+1$ such terms requires total size $\bigO{ds}$.
The top-level summation uses one addition gate and $d+1$ scalar multiplications by the $z_{d + 1 - i, j}$ along $d+1$ wires.
Therefore, \Cref{eq:H_i_explicit} yields a $g$-oracle circuit $D_i$ of size $\bigO{ds+d} = \bigO{ds}$ and depth $\Delta+1$ computing $\coeff_{t^i}(f)$.
Moreover, if the top gate of each $C_{\alpha_j}$ is an addition gate, then the top two layers of $D_i$ consist entirely of additions and can be merged to reduce the depth bound to $\Delta$.
\end{pf}

\begin{remark}
The condition $\abs{\F} \geq d + 1$ in \Cref{lem:homogenization}, required for interpolation, is the sole reason our main theorems assume the base field $\F$ to be sufficiently large. 
We also note that if $g$ is given not as a black box but as a white-box circuit, then its homogeneous components can be extracted in a gate-by-gate fashion~\cite{Str73}, which requires no lower bound on $\abs{\F}$.
\end{remark}

\section{Determinantal Ideals}\label{sec:det}

% Added a two sentence description of where our proof and the AF proof are the same and different. Seems like this was already done for the Pfaffian case.
Our application of the determinantal straightening law  (\Cref{cor:det-basis}), together with the isolation lemma (\Cref{cor:isolation_exponents}) and extraction of coefficients (\Cref{lem:homogenization}), allows us to isolate a canonical bitableau from the expansion of nonzero $f \in I^{\det}_{n, m, r}$ in the basis of standard bideterminants.
In particular, we will show that this can be done via a polynomial-sized depth-three $f$-oracle circuit using the isolation lemma (\Cref{cor:isolation_exponents}) as well as coefficient extraction (\Cref{lem:homogenization}).
This is in contrast to the proof of~\cite{AF21}, where they use a Kronecker-type substitution in order to isolate terms.
To do this, we will give a more careful analysis of the terms that arise when applying the substitution operators to the expression of a polynomial $f(X) \in I^{\det}_{n, m, r}$.

Throughout, fix a field $\F$.
Let $n, m$ be natural numbers, and for $1 \leq i \leq n$ and $1 \leq j \leq m$, let $x_{i, j}$ be indeterminates.
Throughout this section, let $X = (x_{i, j})$ be an $n \times m$ matrix of variables.
Let $1 \leq r \leq \min(n, m)$ and define $I^{\det}_{n, m, r}$ to be the ideal in $\F[X]$ generated by the $r \times r$ minors of $X$.

Our first step will be to establish a series of linear transformations that send a conjugate semistandard bitableau to the anti-canonical conjugate semistandard bitableau of the same shape.
\begin{defn}[Substitution operator]\label{defn:sub}
    Let $i < j$.
    Define the operator $\Sub_{i \to j}$ which takes as input a tableau $S$ and returns the tableau $S'$ which is formed by taking each row of $S$ which has an $i$ but not a $j$, replacing that $i$ with $j$, and then sorting the row in increasing order.
    We also define $h_{i}^{j}(S)$ to be the number of times $i$ is replaced by $j$ after applying $\Sub_{i \to j}$ to $S$.
\end{defn}

While these substitution operators $\Sub_{i \to j}$ on their own are not injective, they are injective when restricted to the set of conjugate semistandard tableau if we also make use of the value of $h_{i}^{j}$.
\begin{lem}[{\cite[Proposition 1.6]{dCEP80}}]\label{lem:sub_injective}
    Let $i, j \in [n]$.
    Consider a conjugate semistandard tableau $S$ such that if a row in $S$ contains an integer $k \leq i$, then that same row in $S$ contains all integers in $\set{i, i + 1, \ldots, j - 1}$.
    Then $\Sub_{i \to j}(S)$ is again conjugate semistandard. 
    Furthermore, $S$ is completely determined by $\Sub_{i \to j}(S)$ and $h_{i}^{j}(S)$.
\end{lem}

We will primarily use \Cref{lem:sub_injective} in the context of successive applications.
In particular, we will be able to completely determine the original tableau from the tableau obtained after successive applications of $\Sub_{i \to j}$ together with the values of $h_i^j$.
We then apply these successive substitutions in the following lexicographic order:
\begin{defn}[Lexicographic ordering]\label{defn:ordering}
    We order the elements of $\binom{[n]}{2}=\set{(i, j) | 1 \leq i < j \leq n}$ such that $(i,j) \preceq (i',j')$ if either $i < i'$, or $i = i'$ and $j \leq j'$.
\end{defn}
In particular, note that the hypothesis in \Cref{lem:sub_injective} holds trivially for the operator $\Sub_{1 \to 2}$.
Furthermore, as we apply the substitution operators successively in the lexicographic ordering, the resulting tableaux also satisfy the hypothesis.

\begin{lem}[{implicit in the proof of~\cite[Corollary 1.7]{dCEP80}; see~\cite[Claim 3.2]{AF21} for a proof}]\label{lem:sub_successive_hyp}
    Fix a partition $\sigma$.
    For a natural number $n$ and $1 \leq i < j \leq n$ and a conjugate semistandard tableau $S$ of shape $\sigma$, define $S_{i', j'}$ where
    \[
        S_{i', j'} = (\Sub_{i' \to j'} \circ \cdots \circ \Sub_{2 \to n}\circ\cdots\circ\Sub_{2 \to 3} \circ \Sub_{1 \to n} \circ \cdots \circ \Sub_{1 \to 2})(S),
    \]
    and $(i', j')$ is the immediate predecessor of $(i, j)$ in the lexicographic ordering on $\binom{[n]}{2}$.
    By convention, if $(i, j) = (1, 2)$, then we  define $S_{i', j'} = S$.
    Then $S_{i', j'}$ satisfies the hypothesis of \Cref{lem:sub_injective} for $\Sub_{i \to j}$, meaning that if a row of $S_{i', j'}$ contains an integer $k \leq i$, then that same row in $S$ contains all integers in $\set{i, i + 1, \ldots, j - 1}$.
\end{lem}

\begin{lem}\label{lem:sub_lex_injective}
    Fix a partition $\sigma$.
    For a natural number $n$ and $1 \leq i < j \leq n$ and a conjugate semistandard tableau $S$ of shape $\sigma$, define $h_{n, i, j}$ as $h_{i}^{j}(S_{i', j'})$ where
    \[
        S_{i', j'} = (\Sub_{i' \to j'} \circ \cdots \circ \Sub_{2 \to n}\circ\cdots\circ\Sub_{2 \to 3} \circ \Sub_{1 \to n} \circ \cdots \circ \Sub_{1 \to 2})(S),
    \]
    and $(i', j')$ is the immediate predecessor of $(i, j)$ in the ordering on $\binom{[n]}{2}$.
    By convention, if $(i, j) = (1, 2)$, then we define $S_{i', j'} = S$.
    Then $S$ is completely determined by $S_{i, j}(S)$ and the ordered sequence $\set{h_{n, a, b}(S)}_{(a, b) \preceq (i, j)}$. 
\end{lem}
\begin{pf}
    We prove via induction on $(i, j)$ in the lexicographic order.
    If $(i, j) = (1, 2)$, this follows from \Cref{lem:sub_injective} as $S$ is completely determined by $\Sub_{1 \to 2}(S)$ and $h_{1}^{2}(S)$.
    
    Now, suppose $(i, j)$ is strictly larger than $(1, 2)$ in the lexicographic ordering, and let $(i', j')$ be the immediate predecessor of $(i, j)$. Assume the claim holds for $(i', j')$ in place of $(i,j)$.
    As $h_{n, i, j}(S) = h_{i}^{j}(S_{i', j'})$, we have that $S_{i', j'}(S)$ is completely determined by $S$ and $h_{n, i, j}(S)$ via \Cref{lem:sub_injective} and \Cref{lem:sub_successive_hyp}.
    By the induction hypothesis, $S$ is completely determined by $S_{i', j'}(S)$ and the sequence $\set{h_{n, a, b}(S)}_{(a, b) \preceq (i', j')}$.
    Thus $S$ is completely determined by $S_{i, j}(S)$ and the sequence $\set{h_{n, a, b}(S)}_{(a, b) \preceq (i, j)}$, completing the proof.
\end{pf}

Applying the substitution operators along the full lexicographic ordering transforms a conjugate semistandard tableau into the anti-canonical tableau of the same shape:
\begin{lem}[{\cite[stated before Corollary 1.7]{dCEP80}}]\label{lem:sub_anticanonical}
    Let $S$ be a conjugate semistandard tableau of shape $\sigma$ such that $S$ has entries of value at most $n$.
    Then
    \[
        (\Sub_{n - 1 \to n} \circ \Sub_{n - 2 \to n} \circ  \Sub_{n - 2 \to n-1} \circ \cdots \circ \Sub_{2 \to n}\circ\cdots\circ\Sub_{2 \to 3} \circ \Sub_{1 \to n} \circ \cdots \circ \Sub_{1 \to 2})(S) = \overline{K}_\sigma.
    \]
\end{lem}

We now study how these substitution operators act on bideterminants $(S \mid T)(X)$.
From properties of the determinant, we see that multiplying the matrix of variables $X$ by a elementary matrix $E_{i, j}$ (\Cref{defn:elem_mat}) corresponds to applying substitution operators on $S$ or $T$:
\begin{lem}\label{lem:Eij_left_mult}
    Let $\lambda$ be a new indeterminate.
    Let $(S \mid T)$ be a bitableau, not necessarily standard, of shape $\sigma$.
    For $0 \leq h \leq h_{i}^{j}(S) - 1$, let $\mathcal{C}^{h}_{i \to j}(S)$ be the set of tableaux of shape $\sigma$ obtained by changing $i$ to $j$ at exactly $h$ rows of $S$ which contain $i$ but not $j$ and reordering those rows to be increasing.
    Then we have that
    \begin{equation}\label{eq:Eij_left_mult}
        (S \mid T)(E_{i, j}(\lambda)X) = \pm \lambda^{h_{i}^{j}(S)} (\Sub_{i \to j}(S) \mid T)(X) + \sum_{h = 0}^{h_{i}^{j}(S) - 1} \lambda^h \sum_{S' \in \mathcal{C}^{h}_{i \to j}(S)} \pm  (S' \mid T)(X).
    \end{equation}
    Similarly, we also have that
    \begin{equation}\label{eq:Eij_right_mult}
        (S \mid T)(X E_{i, j}(\lambda)^{\top}) = \pm \lambda^{h_{i}^{j}(T)} (S \mid \Sub_{i \to j}(T))(X) + \sum_{h = 0}^{h_{i}^{j}(T) - 1} \lambda^h \sum_{T' \in \mathcal{C}^{h}_{i \to j}(T)} \pm (S \mid T')(X).
    \end{equation}
\end{lem}
\begin{pf}
    We only prove \Cref{eq:Eij_left_mult} as the proof of \Cref{eq:Eij_right_mult} is completely analogous.
    First, suppose $\sigma$ is exactly one row.
    Then $(S \mid T)(X)$ is just the determinant of some submatrix of $X$ with rows specified by $S$ and columns specified by $T$.
    Let $\widetilde{X}$ be matrix formed by replacing row $i$ of $X$ with $\lambda$ times row $j$ and let $X' = X + \widetilde{X}$.
    Thus, $E_{i, j}(\lambda)X = X'$.
    More explicitly, $X'$ is formed by adding $\lambda$ times row $j$ to row $i$.
    We will consider cases based on if $S$ contains $i$ or $j$ and make use of the multilinear and alternating properties of the determinant.
    If $S$ does not contain $i$, then
    \[
        (S \mid T)(X') = (S \mid T)(X)
    \]
    because $(S \mid T)$ does not use row $i$ and $X$ and $X'$ differ only in row $i$.
    If $S$ contains both $i$ and $j$, then as the determinant is alternating and the $i$-th row and $j$-th row of $\widetilde{X}$ only differ by a multiple, we must have that $(S \mid T)(\widetilde{X}) = 0$.
    Therefore, by the multilinearity of the determinant we can write
    \[
        (S \mid T)(X') = (S \mid T)(X) + (S \mid T)(\widetilde{X}) = (S \mid T)(X).
    \]
    Finally, if $S$ contains $i$ but not $j$, then by multilinearity of the determinant, we have that
    \[
        (S \mid T)(X') = (S \mid T)(X) + \lambda(S \mid T)(X') = (S \mid T)(X) \pm \lambda(\Sub_{i \to j}(S) \mid T)(X).
    \]
    The $\pm$ in front of the second term comes from the fact that $\Sub_{i \to j}$ not only changes $i$ to $j$ but also reorders the entries to be in increasing order again.
    Reordering rows induces a change in sign that we must account for.
    Thus, we overall have that
    \begin{equation}\label{eq:Eij_left_mult_cases}
        (S \mid T)(E_{i, j}(\lambda)X) = 
            \begin{cases}
                (S \mid T)(X) \pm \lambda(\Sub_{i \to j}(S) \mid T)(X) & \text{if } S \text{ contains } i \text{ but not } j, \\
                (S \mid T)(X) & \text{otherwise}.
            \end{cases}
    \end{equation}
    This gives \Cref{eq:Eij_left_mult} in the case of a single row.
    
    The result for a general bitableau $(S \mid T)$ of shape $\sigma$ then follows via repeated application of the one row case.
    As an application of the definition of a bideterminant (\Cref{defn:bitableau}), we can write $(S \mid T)(X)$ as a product of one row bideterminants:
    \begin{equation}\label{eq:Eij_left_each_row}
        (S \mid T)(E_{i, j}(\lambda)X) = \prod_{k = 1}^{\widehat{\sigma}_1} (S_k \mid T_k)(E_{i, j}(\lambda)X).
    \end{equation}
    We then apply the one row case (\Cref{eq:Eij_left_mult_cases}) to each factor in the product in \Cref{eq:Eij_left_each_row}, expand the product, and study the terms of various degree in $\lambda$ after expansion.
    For each $0 \leq h \leq h_{i}^{j}(S)$, it is clear that terms of degree $h$ in $\lambda$ are of the form $\pm\lambda^{h}(S' \mid T)(X)$ where $S'$ is obtained by changing $i$ to $j$ in exactly $h$ rows of $S$ which contain $i$ but not $j$ and reordering those rows to be increasing.
    Since $\mathcal{C}^{h_{i}^{j}(S)}_{i \to j}(S) = \set{\Sub_{i \to j}(S)}$, there is exactly one term of degree $h_{i}^{j}(S)$ in $\lambda$, namely $\pm \lambda^{h_{i}^{j}(S)} (\Sub_{i \to j}(S) \mid T)(X)$
    This proves \Cref{eq:Eij_left_mult}.
    The proof of \Cref{eq:Eij_right_mult} is completely analogous.
\end{pf}

\subsection{Isolating One Term}\label{sec:extract_det}  

In this subsection, in addition to the variables $x_{i,j}$, where $i\in [n]$ and $j\in [m]$, we introduce new sets of indeterminates $\Lambda = \set{\lambda_{i, j} | 1 \leq i < j \leq n}$ and $\Xi = \set{\xi_{i, j} | 1 \leq i < j \leq m}$.
We have that $\abs{\Lambda} = \binom{n}{2} = \bigO{n^2}$ and $\abs{\Xi} = \binom{m}{2} = \bigO{m^2}$.
The purpose of these new indeterminates is that they will allow us to keep track of the effects of substitution operators as we isolate a canonical bitableau.

Define $M \in \F[\Lambda]^{n\times n}$ and $N \in \F[\Xi]^{m\times m}$ by
\begin{align*}
M ={}& E_{1, 2}(\lambda_{1, 2}) \cdots E_{1, n}(\lambda_{1, n}) E_{2, 3}(\lambda_{2, 3}) \cdots E_{2, n}(\lambda_{2, n}) \\ 
&\cdots E_{n - 2, n - 1}(\lambda_{n - 2, n - 1}) E_{n - 2, n}(\lambda_{n - 2, n}) E_{n - 1, n}(\lambda_{n - 1, n})
\end{align*}
and
\begin{align*}
N ={}& E_{m - 1, m}(\xi_{m - 1, m})^{\top} E_{m - 2, m}(\xi_{m - 2, m})^\top E_{m - 2, m - 1}(\xi_{m - 2, m - 1})^\top \\ &
\cdots E_{2, m}(\xi_{2, m})^{\top} \cdots E_{2, 3}(\xi_{2, 3})^\top E_{1, m}(\xi_{1, m})^\top \cdots E_{1, 2}(\xi_{1, 2})^\top.
\end{align*}

Recall the lexicographic ordering $\preceq$ (\Cref{defn:ordering}) on the elements of $\binom{[n]}{2}$.
This induces a lexicographic order on $(\Z_{\geq 0})^{\binom{[n]}{2}}$.
In the same way, we order the elements of $\binom{[m]}{2} = \set{(i, j) | 1 \leq i < j \leq m}$, which induces a lexicographic order on $(\Z_{\geq 0})^{\binom{[m]}{2}}$.
This also induces a lexicographic term order on the monomials in the variables $\Lambda$ and $\Xi$ via their degree vectors.
Finally, we define the lexicographic order on $(\Z_{\geq 0})^{\binom{[n]}{2}} \times (\Z_{\geq 0})^{\binom{[m]}{2}}$ by comparing the first component before the second component.
We use $M$ and $N$ to obtain polynomials in $\F[X, \Lambda, \Xi]$ with some terms consisting of anti-canonical bideterminants.

\begin{lem}\label{lem:bideterminant_leading_terms}
Let $f(X) \in I^{\det}_{n, m, r}$ be a nonzero polynomial of degree $d$.
Then $f(MX)$ can be expressed as a sum
\begin{equation}\label{eq:expansion1}
    f(MX) = \sum_{k \in A} \widetilde{c}_{k} \Lambda^{\overline{e}_k}\cdot (\overline{K}_{\sigma_k}\mid T_k)(X) + \sum_{\ell\in A'} \widetilde{c}_{\ell} \Lambda^{\overline{e}_\ell}\cdot (S_\ell\mid T_\ell)(X),
\end{equation}
such that the following hold:
\begin{enumerate}
    \item\label{item:expansion1_i0} $A$ is a nonempty finite set and $A'$ is finite set disjoint from $A$.
    \item\label{item:expansion1_i1} All terms in \Cref{eq:expansion1} have the form $c \Lambda^{\overline{e}}\cdot (S\mid T)(X)$, where $c\in \F^\times$, $\overline{e}\in \set{0,1,\dots,d}^{\binom{[n]}{2}}$,
    and $(S\mid T)(X)$ is a bideterminant in $I^{\det}_{n, m, r}$ of degree $d$.
    In particular, if $S$ and $T$ have shape $\sigma$, then $\abs{\sigma} \leq d$.
    \item $T_k$ is a conjugate semistandard tableau of shape $\sigma_k$ for $k\in A$.
    \item\label{item:expansion1_i1.5} For every $\ell\in A'$, there exists $k=k(\ell)\in A$ such that $\overline{e}_\ell$ is strictly smaller than $\overline{e}_k$ in the lexicographic order and the shape $\sigma_\ell$ of $(S_\ell\mid T_\ell)$ equals the shape $\sigma_k$.
    \item\label{item:expansion1_i2} The triples $(\overline{e}_k,\sigma_k,T_k)$ are distinct as $k$ ranges over $A$.
\end{enumerate} 
\end{lem}

\begin{pf}
By \Cref{cor:det-basis} and \Cref{prop:det_shape}, we can expand $f(X)$ as a linear combination of standard bideterminants over $\F$
\[
    f(X)=\sum_{k \in B} c_k \cdot (S_k\mid T_k)(X)
\]
where $B$ is nonempty, each $c_k \in \F^\times$, and $(S_k \mid T_k)(X)$ is a standard bideterminant in $I^{\det}_{n,m,r}$ of shape $\sigma_k$ where $\abs{\sigma_k} \leq d$ for $k \in A$, with these bideterminants being distinct as $k$ ranges over $B$.
It follows that 
\[
    f(MX)=\sum_{k\in B} c_k \cdot (S_k\mid T_k)(MX).
\]
For each $k\in B$, we expand the term $c_k \cdot (S_k\mid T_k)(MX)$ by repeatedly applying \Cref{lem:Eij_left_mult}.
This expansion shows that $c_k \cdot (S_k\mid T_k)(MX)$ equals $\pm c_k \Lambda^{\overline{e}_k}\cdot (\overline{K}_{\sigma_k}\mid T_k)(X)$ with $\overline{e}_k=(h_{n,i,j}(S_k))_{1\leq i<j\leq n} \in \set{0, 1, \dots, d}^{\binom{[n]}{2}}$ plus some terms of the form $\pm \widetilde{c}_{\ell}\Lambda^{\overline{e}_\ell}\cdot (S_\ell\mid T_\ell)(X)$, each with a new index $\ell$,
where $\widetilde{c}_{\ell}=\pm c_k$, $(S_\ell\mid T_\ell)(X)$ is a (not necessarily standard) bideterminant of shape $\sigma_\ell=\sigma_k$, and $\overline{e}_\ell$ is strictly smaller than $\overline{e}_k$ in the lexicographic order.
Note that as $\sigma_\ell=\sigma_k$, their first rows both have length at least $r$, implying that $(S_\ell\mid T_\ell)(X)\in I^{\det}_{n,m,r}$ by \Cref{prop:det_shape}.
Also note that $\overline{e}_\ell \in \set{0, 1, \dots, d}^{\binom{[n]}{2}}$ since a substitution $i\mapsto j$ can be performed at most $(\widehat{\sigma_\ell})_1\leq d$ times to any tableau of shape $\sigma_\ell$.
We add $\ell$ to $A'$ for each such term and let $k(\ell) = k$ and add this $k$ to $A$.

Therefore, we have the expansion
\[
    f(MX) = \sum_{k \in A} \pm{c}_{k} \Lambda^{\overline{e}_k}\cdot (\overline{K}_{\sigma_k}\mid T_k)(X) + \sum_{\ell\in A'} \widetilde{c}_{\ell} \Lambda^{\overline{e}_\ell}\cdot (S_\ell\mid T_\ell)(X)
\]
satisfying \Crefrange{item:expansion1_i0}{item:expansion1_i1.5}.

It remains to prove \Cref{item:expansion1_i2}.
Assume to the contrary that there exist distinct $k_1,k_2\in A$ such that $(\overline{e}_{k_1},\sigma_{k_1},T_{k_1})=(\overline{e}_{k_2},\sigma_{k_2},T_{k_2})$.
As the bideterminants $(S_k\mid T_k)$ are distinct as $k$ ranges over $A$, we have $S_{k_1}\neq S_{k_2}$, both of which are conjugate semistandard.
From the proof above, we have $\overline{e}_{k_1}=(h_{n,i,j}(S_{k_1}))_{1\leq i < j \leq n}$
and $\overline{e}_{k_2}=(h_{n,i,j}(S_{k_2}))_{1\leq i<j\leq n}$.
Note that $S_{k_1}$ and $S_{k_2}$ have the same shape $\sigma_{k_1}=\sigma_{k_2}$.
By \Cref{lem:sub_lex_injective}, $S_{k_1}$ is completely determined by $\overline{K}_{\sigma_{k_1}}$ and $(h_{n,i,j}(S_{k_1}))_{1\leq i<j \leq n}$, and $S_{k_2}$ is completely determined by $\overline{K}_{\sigma_{k_2}}$ and $(h_{n,i,j}(S_{k_2}))_{1\leq i < j \leq n }$.
But then if $\overline{e}_{k_1} = \overline{e}_{k_2}$ and $\sigma_{k_1} = \sigma_{k_2}$, we must have that $S_{k_1} = S_{k_2}$, a contradiction.
\end{pf}

\begin{lem}\label{lem:twosides}
Let $f(X) \in I^{\det}_{n, m, r}$ 
be a nonzero polynomial of degree $d$.
Then $f(MXN)$ can be expressed as a sum
\begin{equation}\label{eq:expansion2}
        f(MXN) = \sum_{k \in A} \widehat{c}_{k} \Lambda^{\overline{e}_k}\Xi^{\overline{f}_k}\cdot (\overline{K}_{\sigma_k}\mid \overline{K}_{\sigma_k})(X) + \sum_{\ell \in B} \widehat{c}_{\ell} \Lambda^{\overline{e}_\ell}\Xi^{\overline{f}_\ell}\cdot (S_\ell\mid T_\ell)(X),
    \end{equation}
such that the following hold:
\begin{enumerate}
    \item\label{item:expansion2_i0} $A$ is a nonempty finite set and $B$ is finite set disjoint from $A$.
    \item\label{item:expansion2_i1} All terms in \Cref{eq:expansion2} have the form $c \Lambda^{\overline{e}}\Xi^{\overline{f}}\cdot (S\mid T)(X)$, where $c\in \F^\times$, $\overline{e} \in \set{0,1,\dots,d}^{\binom{[n]}{2}}$, $\overline{f}\in \set{0,1,\dots,d}^{\binom{[m]}{2}}$,
    and $(S\mid T)(X)$ is a bideterminant in $I^{\det}_{n, m, r}$ of degree at most $d$.
    In particular, if $S$ and $T$ have shape $\sigma$, then $\abs{\sigma} \leq d$ 
    \item\label{item:expansion2_i1.5} For every $\ell\in B$, there exists $k=k(\ell)\in A$ such that $(\overline{e}_\ell, \overline{f}_\ell)$ is strictly smaller than $(\overline{e}_k,\overline{f}_k)$ in the lexicographic order and the shape $\sigma_\ell$ of $(S_\ell\mid T_\ell)$ equals the shape $\sigma_k$.
    \item\label{item:expansion2_i2} The triples $(\overline{e}_k,\overline{f}_k,\sigma_k)$ are distinct, where $k$ ranges over $A$.
\end{enumerate} 
\end{lem}

\begin{pf}
We use \Cref{lem:bideterminant_leading_terms} to write $f(MX)$ as 
\[
    f(MX) = \sum_{k \in A} \widetilde{c}_k \Lambda^{\overline{e}_k}\cdot (\overline{K}_{\sigma_k}\mid T_k)(X) + \sum_{\ell\in A'} \widetilde{c}_{\ell} \Lambda^{\overline{e}_\ell}\cdot (S_\ell\mid T_\ell)(X)
\]
which satisfies \Cref{lem:bideterminant_leading_terms}.
Then
\[
f(MXN)=\sum_{k \in A} \widetilde{c}_{k} \Lambda^{\overline{e}_k}\cdot (\overline{K}_{\sigma_k}\mid T_k)(XN)
+\sum_{\ell \in A'} \widetilde{c}_{\ell} \Lambda^{\overline{e}_\ell}\cdot (S_\ell\mid T_\ell)(XN).
\]

For each $k\in A$, we expand the term $\widetilde{c}_{k} \Lambda^{\overline{e}_k}\cdot (\overline{K}_{\sigma_k}\mid T_k)(XN)$ by repeatedly applying \Cref{lem:Eij_left_mult}.
This expansion shows that $\widetilde{c}_{k} \Lambda^{\overline{e}_k}\cdot (\overline{K}_{\sigma_k}\mid T_k)(XN)$ equals $\pm \widetilde{c}_{k} \Lambda^{\overline{e}_k}\Xi^{\overline{f}_k}\cdot (\overline{K}_{\sigma_k}\mid \overline{K}_{\sigma_k})(X)$ with $\overline{f}_k=(h_{m,i,j}(T_k))_{1\leq i<j\leq m} \in \set{0,1,\dots,d}^{\binom{[m]}{2}}$, 
plus some terms of the form $\widehat{c}_{\ell}\Lambda^{\overline{e}_\ell}\Xi^{\overline{f}_\ell}\cdot (S_\ell\mid T_\ell)(X)$, each with a new index $\ell$,
where $\widehat{c}_{\ell}=\pm \widetilde{c}_k$, $(S_\ell\mid T_\ell)(X)$ is a (not necessarily standard) bideterminant of shape $\sigma_\ell=\sigma_k$, $\overline{e}_\ell=\overline{e}_k$, and $\overline{f}_\ell$ is strictly smaller than $\overline{f}_k$ in the lexicographic order.
In particular, $(\overline{e}_\ell, \overline{f}_\ell)$ is strictly smaller than $(\overline{e}_k, \overline{f}_k)$ in the lexicographic order.
Note that as $\sigma_\ell=\sigma_k$, their first rows both have length at least $r$, implying that $(S_\ell\mid T_\ell)(X)\in I^{\det}_{n,m,r}$ by \Cref{prop:det_shape}.
Also note that $\overline{f}_\ell \in \set{0,1,\dots,d}^{\binom{[m]}{2}}$ since a substitution $i\mapsto j$ can be performed at most $(\widehat{\sigma_\ell})_1 \leq d$ times to any tableau of shape $\sigma_\ell$.
We add $\ell$ to $B$ for each such term and let $k(\ell)=k$.

For each $\ell\in A'$, we expand the term $\widetilde{c}_{\ell} \Lambda^{\overline{e}_\ell}\cdot (S_\ell\mid T_\ell)(XN)$ by repeatedly applying \Cref{lem:Eij_left_mult}.
Note that $(S_\ell\mid T_\ell)$ is not necessarily standard, but \Cref{lem:Eij_left_mult} still applies.
Consider any term $\widehat{c}_{\ell'}\Lambda^{\overline{e}_{\ell'}}\Xi^{\overline{f}_{\ell'}}\cdot (S_{\ell'}\mid T_{\ell'})(X)$ generated this way, where $\ell'$ is a new index.
Let $k=k(\ell)$ as in \Cref{lem:bideterminant_leading_terms}~\eqref{item:expansion1_i1.5}.
We have $\widehat{c}_{\ell'}=\pm\widetilde{c}_{\ell}$, $(S_{\ell'}\mid T_{\ell'})(X)$ is a (not necessarily standard) bideterminant of shape $\sigma_{\ell'}=\sigma_\ell=\sigma_k$, and $\overline{e}_{\ell'}=\overline{e}_\ell$, which is strictly smaller than $\overline{e}_k$ in the lexicographic order.
In particular, $(\overline{e}_{\ell'}, \overline{f}_{\ell'})$ is strictly smaller than $(\overline{e}_k, \overline{f}_k)$ in the lexicographic order.
Note that as $\sigma_{\ell'}=\sigma_k$, their first rows both have length at least $r$, implying that $(S_{\ell'}\mid T_{\ell'})(X)\in I^{\det}_{n,m,r}$ by \Cref{prop:det_shape}.
Also note that $\overline{f}_{\ell'} \in \set{0,1,\dots,d}^{\binom{[m]}{2}}$ for the same reason as in the previous paragraph.
We add $\ell'$ to $B$ for each such term and let $k(\ell')=k$.

Therefore, we have the expansion
\[
    f(MXN) = \sum_{k \in A} \pm\widetilde{c}_{k} \Lambda^{\overline{e}_k}\Xi^{\overline{f}_k}\cdot (\overline{K}_{\sigma_k}\mid \overline{K}_{\sigma_k})(X) + \sum_{\ell \in B} \widehat{c}_{\ell} \Lambda^{\overline{e}_\ell}\Xi^{\overline{f}_\ell}\cdot (S_\ell\mid T_\ell)(X),
\]
satisfying \Crefrange{item:expansion2_i0}{item:expansion2_i1.5}.

It remains to prove \Cref{item:expansion2_i2}.
Assume to the contrary that there exist distinct $k_1,k_2\in A$ such that $(\overline{e}_{k_1},\overline{f}_{k_1},\sigma_{k_1})=(\overline{e}_{k_2},\overline{f}_{k_2},\sigma_{k_2})$.
By \Cref{lem:bideterminant_leading_terms}~\eqref{item:expansion1_i2}, we have $T_{k_1}\neq T_{k_2}$, both of which are conjugate semistandard.
From the proof above, we have $\overline{f}_{k_1}=(h_{m,i,j}(T_{k_1}))_{1\leq i<j\leq m}$ and $\overline{f}_{k_2}=(h_{m,i,j}(T_{k_2}))_{1\leq i<j\leq m}$.
Note that $T_{k_1}$ and $T_{k_2}$ have the same shape $\sigma_{k_1}=\sigma_{k_2}$.
By \Cref{lem:sub_lex_injective}, $T_{k_1}$ is completely determined by $\overline{K}_{\sigma_{k_1}}$ and $(h_{m,i,j}(T_{k_1}))_{1\leq i<j\leq m}$, and $T_{k_2}$ is completely determined by $\overline{K}_{\sigma_{k_2}}$ and $(h_{m,i,j}(T_{k_2}))_{1\leq i<j\leq m}$.
But then if $\overline{f}_{k_1} = \overline{f}_{k_2}$ and $\sigma_{k_1} = \sigma_{k_2}$, we must have that $T_{k_1} = T_{k_2}$, a contradiction.
\end{pf}

Next, we reformulate \Cref{lem:twosides} by permuting the rows and columns of $X$, thereby transforming the anti-canonical bitableaux into canonical tableaux.
This transformation is not essential, but it simplifies the notation and analysis.

For an integer $k>0$, let $J_{k}\in\F^{k\times k}$ be the $k\times k$ matrix with 1’s along the anti-diagonal from the bottom left to the top right and 0’s elsewhere.
Note that $\det_{k}(J_k) = (-1)^{\binom{k}{2}} = \pm 1$.
The substitution $X\mapsto J_{n} X J_{m}$ has the effect of replacing row $i$ with row $n-i+1$ and column $j$ with column $m-j+1$.
This turns an anti-canonical bitableau $(\overline{K}_\sigma\mid \overline{K}_\sigma)$ into a canonical one $(K_\sigma\mid K_\sigma)$.
\begin{cor}\label{cor:reformuated}
    Let $f(X) \in I^{\det}_{n, m, r}$ 
    be a nonzero polynomial of degree $d$.
    Then $f(M J_n X J_m N)$ can be expressed as a sum
    \begin{equation}\label{eq:expansion3}
            f(M J_n X J_m N) = \sum_{k \in A} \widehat{c}_{k} \Lambda^{\overline{e}_k}\Xi^{\overline{f}_k}\cdot (K_{\sigma_k}\mid K_{\sigma_k})(X) + \sum_{\ell \in B} \widehat{c}_{\ell} \Lambda^{\overline{e}_\ell}\Xi^{\overline{f}_\ell}\cdot (S_\ell\mid T_\ell)(X),
        \end{equation}
    such that the following hold:
    \begin{enumerate}
        \item\label{item:expansion3_i0} $A$ is a nonempty finite set and $B$ is finite set disjoint from $A$.
        \item\label{item:expansion3_i1} All terms in \Cref{eq:expansion3} have the form $c \Lambda^{\overline{e}}\Xi^{\overline{f}}\cdot (S\mid T)(X)$, where $c\in \F^\times$, $\overline{e}\in \set{0,1,\dots,d }^{\binom{[n]}{2}}$, $\overline{f}\in \set{0,1,\dots,d}^{\binom{[m]}{2}}$,
        and $(S\mid T)(X)$ is a bideterminant in $I^{\det}_{n, m, r}$ of degree $d$.
        In particular, if $S$ and $T$ have shape $\sigma$, then $\abs{\sigma} \leq d$.
        \item\label{item:expansion3_i1.5} For every $\ell\in B$, there exists $k=k(\ell)\in A$ such that $(\overline{e}_\ell, \overline{f}_\ell)$ is strictly smaller than $(\overline{e}_k,\overline{f}_k)$ in the lexicographic order and the shape $\sigma_\ell$ of $(S_\ell\mid T_\ell)$ equals the shape $\sigma_k$.
        \item\label{item:expansion3_i2} The triples $(\overline{e}_k,\overline{f}_k,\sigma_k)$ are distinct, where $k$ ranges over $A$.
    \end{enumerate} 
\end{cor}

Next, we introduce new variables $Y = \set{y_1,\dots,y_n}$ and $Z = \set{z_1,\dots,z_m}$.
Define the diagonal matrices 
\[
    D = \mathrm{diag}(y_1,\dots,y_n)\in\F[Y]^{n\times n} \quad\text{and}\quad D'=\mathrm{diag}(z_1,\dots,z_m)\in\F[Z]^{m\times m}.
\]

\begin{lem}\label{lem:diagonals}
Let $(S\mid T)(X)$ be a bideterminant of degree at most $d$.
Then 
\[
  (S\mid T)(D X D')=Y^{\overline{s}} Z^{\overline{t}}\cdot  (S\mid T)(X),
\]
where $\overline{s}=(s_1,\dots,s_n)\in \set{0,1,\dots,d}^n$, $\overline{t}=(t_1,\dots,t_m)\in \set{0,1,\dots,d}^m$, $s_i$ is the number of times $i$ appears in $S$, $t_j$ is the number of times $j$ appears in $T$, \ie, $(s_1 e_1+ \dots + s_n e_n)\oplus (t_1 e_1 + \dots + t_m e_m)$ is the multidegree of $(S\mid T)(X)$ with respect to the grading defined in \Cref{defn:multideg}.
\end{lem}
\begin{pf}
Reduce to the case in which $S$ and $T$ each have one row, and then use the multilinearity of the determinant.
\end{pf}
We will often simply say that $(S \mid T)(X)$ in \Cref{lem:diagonals} has multidegree $(\overline{s}, \overline{t})$ rather than writing it as $(s_1 e_1 + \dots + s_n e_n) \oplus (t_1 e_1 + \dots + t_m e_m)$.

Next, we introduce another variable $v$, and modify $D$ and $D'$ as follows:
\[
    \widetilde{D} = \mathrm{diag}(y_1 v, y_2 v^2\dots,y_n v^n )\in\F[Y,v]^{n\times n} \quad\text{and}\quad \widetilde{D}'=\mathrm{diag}(z_1 v,z_2 v^2,\dots,z_m v^m)\in\F[Z,v]^{m\times m}.
\]
\begin{lem}\label{lem:multiply-by-diag}
    Let $(S\mid T)(X)$ be a bideterminant.
    Then 
    \[
        (S\mid T)(\widetilde{D} X \widetilde{D}')=Y^{\overline{s}} Z^{\overline{t}} v^{\abs{S} + \abs{T}} \cdot (S\mid T)(X),
    \]
    where $(\overline{s},\overline{t})$ is the multidegree of $(S\mid T)(X)$.
\end{lem}

\begin{pf}
Replacing $y_i$ by $y_i v^i$ in $D$ multiplies the expression by $v^i$ exactly $s_i$ times.  
Similarly, replacing $z_j$ by $z_j v^j$ in $D'$ multiplies the expression by $v^j$ exactly $t_j$ times.  
Taking all $i \in [n]$ and $j \in [m]$ into account, the total exponent of $v$ is $\abs{S} + \abs{T}$, \ie, the sum of the entries of $S$ and of $T$.
\end{pf}

Before proceeding, we clarify the relationship between our arguments and those in \cite{AF21}.
Much of our proof follows the overall structure of \cite{AF21}: the expansions in \eqref{eq:expansion1}, \eqref{eq:expansion2}, and \eqref{eq:expansion3} already appear there.
However, our statements refine those in \cite{AF21} by providing more detailed information about the ``garbage'' terms (those indexed by the sets $A'$ or $B$ in \eqref{eq:expansion1}, \eqref{eq:expansion2}, and \eqref{eq:expansion3}). In particular, we establish additional properties of these terms that will be essential in the proof of the key technical lemma below, which serves as preparation for our use of the isolation lemma. 

The main difference between the two approaches lies in how a term associated with a canonical standard bideterminant is singled out.
The proof in \cite{AF21} uses a Kronecker-type substitution and therefore only needs to ensure that the lexicographically leading term is unique.
In contrast, our argument applies the isolation lemma, which isolates a term but does not, \apriori, guarantee that the isolated term corresponds to a canonical standard bideterminant.
The following lemma shows, through a careful analysis that relies on the properties established in \Cref{cor:reformuated}, that isolating a term of lowest degree in $v$ indeed yields a canonical standard bideterminant.

\begin{lem}\label{lem:key}
Let $f(X) \in I^{\det}_{n, m, r}$ be a nonzero polynomial of degree $d$.
Let $g = f(M J_n \widetilde{D} X \widetilde{D}' J_m N) \in \F[X, \Lambda, \Xi, Y, Z, v]$.
View $g$ as a univariate polynomial in $v$ with coefficients in $\F[X,\Lambda,\Xi,Y,Z]$, and write $g=\sum_{i} \coeff_{v^i}(g) v^i$, where $\coeff_{v^i}(g)\in \F[X,\Lambda,\Xi,Y,Z]$ denotes the coefficient of $v^i$ in $g$.
Choose the smallest integer $d_{\min}$ such that  $\coeff_{v^{d_{\min}}}(g)\neq 0$.
Then $d_{\min}=2 \abs{K_\sigma}$ for some shape $\sigma$ with $\abs{\sigma} \leq d$ and we may write $\coeff_{v^{d_{\min}}}(g)$ as a finite sum
\begin{equation}\label{eq:required}
    \coeff_{v^{d_{\min}}}(g)=\sum_{k\in I} c_k \Lambda^{\overline{e}_k}\Xi^{\overline{f}_k}Y^{\overline{s}_k}Z^{\overline{t}_k}\cdot (K_{\sigma_k}\mid K_{\sigma_k})(X), 
\end{equation}
such that the following hold:
\begin{enumerate}
    \item\label{item:required1} For each $k\in I$, $c_k\in\F^\times$,  $(\sigma_k)_1\geq r$, and $\abs{\sigma_k} \leq d$.
    \item\label{item:required2} The tuples $(\overline{e}_k,\overline{f}_k, \overline{s}_k,\overline{t}_k)$ are distinct, where $k$ ranges over $I$.
\end{enumerate}
\end{lem}

\begin{pf}
Consider the expansion
\[
        f(M J_n X J_m N) = \sum_{k \in A} \widehat{c}_{k} \Lambda^{\overline{e}_k}\Xi^{\overline{f}_k}\cdot (K_{\sigma_k}\mid K_{\sigma_k})(X) + \sum_{\ell \in B} \widehat{c}_{\ell} \Lambda^{\overline{e}_\ell}\Xi^{\overline{f}_\ell}\cdot (S_\ell\mid T_\ell)(X)
\]
given by \Cref{cor:reformuated}.
Then we have
\begin{equation}\label{eq:expansion-g}
\begin{aligned}
        g &= \sum_{k \in A} \widehat{c}_{k} \Lambda^{\overline{e}_k}\Xi^{\overline{f}_k}\cdot (K_{\sigma_k}\mid K_{\sigma_k})(\widetilde{D}X\widetilde{D}') + \sum_{\ell \in B} \widehat{c}_{\ell} \Lambda^{\overline{e}_\ell}\Xi^{\overline{f}_\ell}\cdot (S_\ell\mid T_\ell)(\widetilde{D}X\widetilde{D}')\\
        &=\sum_{k \in A} \widehat{c}_{k} \Lambda^{\overline{e}_k}\Xi^{\overline{f}_k}Y^{\overline{s}_k}Z^{\overline{t}_k}v^{2\abs{K_{\sigma_k}}}\cdot (K_{\sigma_k}\mid K_{\sigma_k})(X) + \sum_{\ell \in B} \widehat{c}_{\ell} \Lambda^{\overline{e}_\ell}\Xi^{\overline{f}_\ell}Y^{\overline{s}_\ell}Z^{\overline{t}_\ell}v^{\abs{S_\ell} + \abs{T_\ell}} \cdot (S_\ell\mid T_\ell)(X),
\end{aligned}
\end{equation}
where for $k\in A$, $(\overline{s}_k,\overline{t}_k)$ is the multidegree of $(K_{\sigma_k}\mid K_{\sigma_k})(X)$, and for $\ell \in B$, $(\overline{s}_\ell,\overline{t}_\ell)$ is the multidegree of $(S_\ell\mid T_\ell)(X)$.
Here, the second equality holds by \Cref{lem:multiply-by-diag}.

Choose $k_0 \in A$ so that $\abs{K_{\sigma_{k_0}}}$ is minimized, and subject to this, $(\overline{e}_{k_0}, \overline{f}_{k_0})$ is maximized in the lexicographic order.
Let $d_0 = 2 \abs{K_{\sigma_{k_0}}}$.
We will show that $\coeff_{v^{d_0}}(g)\neq 0$.

For any $k\in A$, consider the term $\widehat{c}_{k} \Lambda^{\overline{e}_k}\Xi^{\overline{f}_k}Y^{\overline{s}_k}Z^{\overline{t}_k}v^{2\abs{K_{\sigma_k}}}\cdot (K_{\sigma_k}\mid K_{\sigma_k})(X)$ in \Cref{eq:expansion-g}.
It is homogeneous of degree $2 \abs{K_{\sigma_k}} \geq 2 \abs{K_{\sigma_{k_0}}} = d_0$ with respect to $v$ by the minimality of $\abs{K_{\sigma_{k_0}}}$, and contributes to $\coeff_{v^{d_0}}(g)$ if and only if $\abs{K_{\sigma_k}} = \abs{K_{\sigma_{k_0}}}$.
Even if it can potentially contribute to $\coeff_{v^{d_0}}(g)$, its (multi)degree $(\overline{e}_k,\overline{f}_k,\overline{s}_k,\overline{t}_k)$ in the variables $\Lambda$, $\Xi$, $Y$, and $Z$ is different from that of the term indexed by $k_0$ unless $k=k_0$.
To see this, suppose $k\neq k_0$.
Then $(\overline{e}_k,\overline{f}_k,\sigma_k)\neq (\overline{e}_{k_0},\overline{f}_{k_0},\sigma_{k_0})$.
If $\sigma_k=\sigma_{k_0}$, then $(\overline{e}_k,\overline{f}_k)\neq (\overline{e}_{k_0},\overline{f}_{k_0})$.
On the other hand, if $\sigma_k\neq \sigma_{k_0}$, then $(\overline{s}_k,\overline{t}_k)\neq (\overline{s}_{k_0},\overline{t}_{k_0})$ by \Cref{lem:canonical_degs}.
Overall, we always have that 
\[
    (\overline{e}_k,\overline{f}_k,\overline{s}_k,\overline{t}_k)\neq (\overline{e}_{k_0},\overline{f}_{k_0},\overline{s}_{k_0},\overline{t}_{k_0}).
\]

Now consider $\ell\in B$ and the term $\mathfrak{m}_\ell\defeq\widehat{c}_{\ell} \Lambda^{\overline{e}_\ell}\Xi^{\overline{f}_\ell}Y^{\overline{s}_\ell}Z^{\overline{t}_\ell}v^{\abs{S_\ell} + \abs{T_\ell}}\cdot (S_\ell\mid T_\ell)(X)$ in \Cref{eq:expansion-g}.
The degree of $\mathfrak{m}_\ell$ in $v$ is $\abs{S_\ell} + \abs{T_\ell}$.
Let $k=k(\ell)$ as in \Cref{cor:reformuated}.
Then by \Cref{cor:reformuated}~\eqref{item:expansion3_i1.5}, $S_\ell$ and $T_\ell$ have shape $\sigma_\ell=\sigma_k$.
Note that among all bideterminants $(S\mid T)$ of shape $\sigma$, the quantity $\abs{S} + \abs{T}$ is minimized when, and only when, $(S\mid T)=(K_\sigma\mid K_\sigma)$.
Thus, we have 
\[
    \abs{S_\ell} + \abs{T_\ell} \geq 2 \abs{K_{\sigma_k}} \geq  2\abs{K_{\sigma_{k_0}}}=d_0
\]
and $\deg_v(\mathfrak{m}_\ell) = \abs{S_\ell} + \abs{T_\ell} = d_0$ if and only if $(S_\ell\mid T_\ell)=(K_{\sigma_k}\mid K_{\sigma_k})$ and $\abs{K_{\sigma_{k}}} = \abs{K_{\sigma_{k_0}}}$.
Even though this can happen, when it happens we have that $(\overline{e}_k,\overline{f}_k)$ is smaller than or equal to $(\overline{e}_{k_0},\overline{f}_{k_0})$ in the lexicographic order by the choice of $k_0$, noting that $\abs{K_{\sigma_{k}}} = \abs{K_{\sigma_{k_0}}}$.
And by \Cref{cor:reformuated}~\eqref{item:expansion3_i1.5}, $(\overline{e}_\ell, \overline{f}_\ell)$ is strictly smaller than $(\overline{e}_k,\overline{f}_k)$ in the lexicographic order.
Overall, either $\deg_v(\mathfrak{m}_\ell)>d_0$, or $(\overline{e}_\ell,\overline{f}_\ell,\overline{s}_\ell,\overline{t}_\ell)\neq (\overline{e}_{k_0},\overline{f}_{k_0},\overline{s}_{k_0},\overline{t}_{k_0})$.

By the above analysis, the term $\widehat{c}_{k_0} \Lambda^{\overline{e}_{k_0}}\Xi^{\overline{f}_{k_0}}Y^{\overline{s}_{k_0}}Z^{\overline{t}_{k_0}}v^{2 \abs{K_{\sigma_{k_0}}}}\cdot (K_{\sigma_{k_0}}\mid K_{\sigma_{k_0}})(X)$ is not canceled by other terms contributing to $\coeff_{v^{d_0}}(g)$ due to the uniqueness of its (multi)degree $(\overline{e}_{k_0},\overline{f}_{k_0},\overline{s}_{k_0},\overline{t}_{k_0})$ in $\Lambda$, $\Xi$, $Y$, and $Z$.
Thus $\coeff_{v^{d_0}}(g)\neq 0$.
The above analysis also shows that $\coeff_{v^{i}}(g)=0$ for $i<d_0$.
Thus, $d_0=d_{\min}$.

Moreover, the above analysis shows that any term that contributes to $\coeff_{v^{d_0}}(g)$ contains a bideterminant of the form $(K_\sigma \mid K_\sigma)(X)$.

We then merge the terms contributing to $\coeff_{v^{d_0}}(g)$ with the same (multi)degree in $\Lambda$, $\Xi$, $Y$, and $Z$.
When we merge two terms $\widehat{c}_{k} \Lambda^{\overline{e}_k}\Xi^{\overline{f}_k}Y^{\overline{s}_k}Z^{\overline{t}_k}v^{d_0}\cdot (K_{\sigma_k}\mid K_{\sigma_k})(X)$ and $\widehat{c}_{k'} \Lambda^{\overline{e}_{k'}}\Xi^{\overline{f}_{k'}}Y^{\overline{s}_{k'}}Z^{\overline{t}_{k'}}v^{d_0}\cdot (K_{\sigma_{k'}}\mid K_{\sigma_{k'}})(X)$
with $(\overline{e}_k,\overline{f}_k,\overline{s}_k,\overline{t}_k)=(\overline{e}_{k'},\overline{f}_{k'},\overline{s}_{k'},\overline{t}_{k'})$,
we always have $(K_{\sigma_{k}}\mid K_{\sigma_{k}})=(K_{\sigma_{k'}}\mid K_{\sigma_{k'}})$.
This follows from \Cref{lem:canonical_degs} and the fact that $(\overline{s}_{k},\overline{t}_{k})$ is the multidegree of $(K_{\sigma_{k}}\mid K_{\sigma_{k}})(X)$ and $(\overline{s}_{k'},\overline{t}_{k'})$ is the multidegree of $(K_{\sigma_{k'}}\mid K_{\sigma_{k'}})(X)$.
Thus, merging the two terms yields a scalar multiple of both.

After merging/canceling terms, we can write $\coeff_{v^{d_{\min}}}(g)=\coeff_{v^{d_0}}(g)$ in the form of \Cref{eq:required} such that \Cref{item:required2} holds.
Furthermore, \Cref{item:required1} holds as well by \Cref{cor:reformuated}~\eqref{item:expansion3_i1} and \Cref{prop:det_shape}.
\end{pf}

\Cref{lem:key} helps us separate a collection of terms solely containing canonical bideterminants.
The next lemma further applies the isolation lemma to single out one term from this collection.

\begin{lem}\label{lem:univariate}
Let $f(X) \in I^{\det}_{n, m, r}$ 
be a nonzero polynomial of degree $d$.
Let $g=f(M J_n \widetilde{D} X \widetilde{D}' J_m N) \in \F[X, \Lambda, \Xi, Y, Z, v]$.
Then there exist integers $z_t = \bigO{d(n^2+m^2)}$ for each variable $t\in \Lambda\sqcup\Xi\sqcup Y\sqcup Z$, and an integer $z_v = \bigO{d^2(n^2+m^2)}$ such that for the two variable substitution maps 
\[
    \phi\colon t \mapsto w^{z_t} \quad\text{ for } t\in \Lambda \sqcup \Xi \sqcup Y \sqcup Z
\]
and
\[
    \psi\colon v \mapsto w^{z_v},
\]
we have that $h \defeq (\psi\circ\phi)(g)\in\F[X,w]$ has $\deg_w(h) = \bigO{d^3(n^3+m^3)}$, and that $\coeff_{w^i}(h) = c\cdot (K_\sigma\mid K_\sigma)(X)$ for some integer $i\leq \deg_w(h)$, where $c\in\F^\times$, $\sigma_1\geq r$ and $\abs{\sigma} \leq d$.
\end{lem}
\begin{pf}
Consider the finite sum from \Cref{lem:key}:
\[
\coeff_{v^{d_{\min}}}(g)=\sum_{k\in I} c_k \Lambda^{\overline{e}_k}\Xi^{\overline{f}_k}Y^{\overline{s}_k}Z^{\overline{t}_k}\cdot (K_{\sigma_k}\mid K_{\sigma_k})(X).
\]
The coordinates of $\overline{e}_k$, $\overline{f}_k$, $\overline{s}_k$, and $\overline{t}_k$ are in $\set{0,1,\dots,d}$ for each $k \in I$ by \Cref{cor:reformuated}~\eqref{item:expansion3_i1} and \Cref{lem:multiply-by-diag}.
Choose a sufficiently large $L=\bigTheta{d(n^2+m^2)}$, and pick integers $z_t$ independently and uniformly at random from $ \set{0, 1, \ldots, L}$, where $t\in \Lambda\sqcup\Xi\sqcup Y\sqcup Z$.
Then by \Cref{cor:isolation_exponents}, with high probability, there exists an integer $d'$ such that the coefficient of the monomial $v^{d_{\min}}w^{d'}$ in $\phi(g)\in\F[X,v,w]=\F[X][v,w]$, has the form $c\cdot (K_\sigma\mid K_\sigma)(X) \in \F[X]$, where $c\in \F^\times$, $\sigma_1\geq r$, and $\abs{\sigma} \leq d$.
Fix the integers $z_t$ such that this occurs.

By \Cref{lem:multiply-by-diag}, the degree of $\phi(g)$ in $v$ is at most $\bigO{d(n+m)}$.
To see this, note that if a tableau of shape $\sigma$ has $\abs{\sigma}\leq d$ and all entries are at most $n$, then the sum of its entries is at most $dn$; similarly, with entries at most $m$ the sum is at most $dm$.
The degree of $\phi(g)$ in $w$ is bounded by $\bigO{dL} = \bigO{d^2(n^2+m^2)}$.
Choose $z_v = \deg_w(\phi(g))+1 = \bigO{d^2(n^2+m^2)}$.
Then the map $\psi: v\mapsto w^{z_v}$ sends the monomials of $\phi(g)$ bijectively to that of $(\psi\circ\phi)(g)=h$, preserving coefficients.
Thus, there exists an integer $i$ such that $\coeff_{w^i}(h)=c\cdot (K_\sigma\mid K_\sigma)$.
Finally, we have that
\[
    \deg_w(h) \leq z_v \cdot \bigO{d(n+m)} + d' \leq \bigO{d^2 (n^2 + m^2)} \cdot \bigO{d(n + m)} + \bigO{d^2(n^2+m^2)} = \bigO{d^3(n^3 + m^3)}.
\]
\end{pf}

We are now ready to prove the main theorem in this subsection.

\begin{thrm}\label{thrm:bideterminant_canonical_extraction}
    Let $f(X) \in I^{\det}_{n, m, r}$ be a nonzero polynomial of degree $d$.
    Assume $\abs{\F} \geq c_0 d^3(n^3+m^3)$, where $c_0>0$ is a large enough constant.
    Then, there exists a depth-three $f$-oracle circuit of size $\bigO{n^2m^2d^3(n^3+m^3)} = \poly(n,m,d)$ computing $(K_\sigma \mid K_\sigma)(X)$, where $\sigma_1 \geq r$ and $\abs{\sigma} \leq d$.
    Furthermore, the top gate of this circuit is an addition gate, and the bottom layer consists of $\bigO{nmd^3(n^3+m^3)}$ addition gates.
    The total number of gates and wires, excluding those between the bottom addition gates and the input gates, is $\bigO{nmd^3(n^3+m^3)}$.\footnote{In the final circuit construction, these wires will be removed and replaced by wires connecting to fewer input gates (see \Cref{thrm:ABP_det_oracle}).
    Hence we state the circuit size bound excluding these wires, which leads to a sharper bound for the final circuit.}
\end{thrm}
\begin{pf}
    Let $h$ be as in \Cref{lem:univariate}.
    Let $\alpha\in\F$ be arbitrary.
    By construction, $h(X,\alpha)$ equals $f(M_\alpha X N_\alpha)$, where $M_\alpha$ (resp. $N_\alpha$) is obtained from $M J_n \widetilde{D}$ (resp. $\widetilde{D}' J_m N$) by substituting powers of $\alpha$ for the variables $\Lambda$, $\Xi$, $Y$, $Z$, and $v$.
    Specifically, if under $\psi \circ \phi$ a variable is mapped to $w^s$ for some integer $s \geq 0$, then we substitute $\alpha^s$ for that variable here.
    As the map $X\mapsto M_\alpha X N_\alpha$ is a linear map, we can construct a depth-two $f$-oracle circuit $C_\alpha$ computing $h(X,\alpha)=f(M_\alpha X N_\alpha)$ as follows:
    Use $nm$ addition gates and $\bigO{n^2m^2}$ wires at the bottom layer computing the $nm$ entries of $M_\alpha X N_\alpha$.
    The top gate is an $f$-gate connecting to the $nm$ addition gates via $nm$ wires.
    Thus, the size of $C_\alpha$ is $\bigO{n^2m^2}$.
    
    Let $d_w=\deg_w(h)$.
    By \Cref{lem:univariate}, we have that $d_w = \bigO{d^3(n^3+m^3)}$ and that $\coeff_{w^i}(h) = c\cdot {(K_\sigma\mid K_\sigma)}(X)$ for some integer $i\leq d_w$, where $c\in\F^\times$, $\sigma_1\geq r$ and $\abs{\sigma} \leq d$.
    As $\abs{\F} \geq c_0 d^3(n^3+m^3)$, where $c_0>0$ is a large enough constant, we may assume $\abs{\F} \geq d_w+1$.
    Then by \Cref{lem:homogenization}, we have a depth-three $f$-oracle circuit $C$ of size $\bigO{d_w(n^2m^2)} = \bigO{n^2m^2d^3(n^3+m^3)}$ computing $\coeff_{w^i}(h)=c\cdot (K_\sigma\mid K_\sigma)$. 
    The top gate of $C$ is an addition gate, which connects to $d_w+1 = \bigO{d^3(n^3+m^3)}$ $f$-gates on the middle layer.
    And these $f$-gates connect to $(d_w+1)nm = \bigO{nmd^3(n^3+m^3)}$ addition gates on the bottom layer.
    
    The above circuit $C$ computes $c \cdot (K_\sigma\mid K_\sigma)(X)$.
    By dividing the multipliers on the wires connecting the top addition gate by $c$, we can transform $C$ into an $f$-oracle circuit with the same underlying graph that computes $(K_\sigma\mid K_\sigma)(X)$.
    This proves the theorem.
\end{pf}

\subsection{Expressing Algebraic Branching Programs as Determinants}

The remainder of the proof follows~\cite{AF21}, which we include here for completeness.
One key idea is exploiting the close connection between two models: algebraic branching programs and determinants.
In particular, we will need the following lemma:

\begin{lem}[{\cite[Lemma 3.6]{AF21},~\cite[Theorem 1]{Val79}}]\label{lem:ABP_to_det}
  Let $g(\overline{y}) \in \F[\overline{y}]$ and suppose $g$ can be computed by an algebraic branching program on $r$ vertices.
  Then there is an $r \times r$ matrix $A$ over $\F[\overline{y}]$ whose entries are polynomials in $\overline{y}$ of degree at most $1$ such that
  \begin{itemize}
    \item $\det_r(A) = 1 + g(\overline{y})$, and
    \item for all $1 \leq k < r$, $\det_k(A_{[k], [k]}) = 1$.
  \end{itemize}
\end{lem}

Algebraic branching programs are closed under homogenization, as stated in the following lemma.

\begin{lem}[{\cite[Lemma 3.7]{AF21}}]\label{lem:homog_ABP}
  Let $g(\overline{y}) \in \F[\overline{y}]$ and suppose $g$ can be computed by an algebraic branching program on $m$ vertices.
  Let $z$ be a new indeterminate.
  Then there is a homogeneous polynomial $\widehat{g}(\overline{y}, z)$ such that $g = \widehat{g}(\overline{y}, 1)$ and $\widehat{g}$ can be computed by an algebraic branching program on $m$ vertices.
\end{lem}

\subsection{Computing the Determinant of a Matrix}

We will use \Cref{thrm:bideterminant_canonical_extraction} to construct a constant-depth algebraic circuit of size $\poly(n,\deg(f))$ with $f$-oracle gates that computes the determinant, thereby resolving \Cref{question:main}. We start by proving the following more general theorem.
\begin{thrm}\label{thrm:ABP_det_oracle}
    Let $f(X) \in I^{\det}_{n, m, r}$ be a nonzero polynomial of degree $d$.
    Let $g(y_1, \ldots, y_\ell) \in \F[\overline{y}]$ be a polynomial computable by an algebraic branching program with at most $r$ vertices.
    Assume $\abs{\F} \geq c_0 d^3(n^3+m^3)$, where $c_0>0$ is a large enough constant.
    Then there is a depth-three $f$-oracle circuit $\Phi$ of size $\bigO{nmd^4\ell r(n^3+m^3)}$ defined over $\F$ such that 
    \begin{itemize}
        \item if $\characteristic(\F) = 0$ then $\Phi$ computes $g(\overline{y})$, and
        \item if $\characteristic(\F) = p > 0$ then $\Phi$ computes $g(\overline{y})^{p^k}$ for some $k \leq \floor{\log_p(d)}$.
    \end{itemize}
    Furthermore, the top layer of this circuit consists of an addition gate.
\end{thrm}
\begin{pf}
    By \Cref{lem:homog_ABP}, there is a homogeneous polynomial $\widehat{g}(\overline{y}, z) \in \F[\overline{y}, z]$ such that $\widehat{g}(\overline{y}, 1) = g(\overline{y})$ and $\widehat{g}(\overline{y}, z)$ can be computed by an algebraic branching program over $\F$ with at most $r$ vertices.
    We may add isolated vertices such that the algebraic branching program has exactly $r$ vertices.
    Then by \Cref{lem:ABP_to_det}, there is a matrix $A(\overline{y}, z) \in \F[\overline{y}, z]^{r \times r}$  such that
    \begin{itemize}
        \item $\det_r(A(\overline{y}, z)) = 1 + \widehat{g}(\overline{y}, z)$,
        \item $\det_i(A(\overline{y}, z)_{[i], [i]}) = 1$ for all $1 \leq i \leq r - 1$,
    \end{itemize}
    and all entries of $A(\overline{y}, z)$ have degree at most $1$ in $\overline{y}$ and $z$.  
    Extend $A(\overline{y}, z)$ to an $n\times m$ matrix by adding $1$'s to the main diagonal and $0$'s elsewhere.

   By \Cref{thrm:bideterminant_canonical_extraction}, there exists a depth-three $f$-oracle circuit $C$ over $\F$ of size $\bigO{n^2m^2d^3(n^3+m^3)}$ computing $(K_\sigma \mid K_\sigma)(X)$ such that $\sigma_1 \geq r$ and $\abs{\sigma} \leq d$.
   Furthermore, the top gate of $C$ is an addition gate, the bottom layer of $C$ consists of $\bigO{nmd^3(n^3+m^3)}$ addition gates, and the total number of gates and wires excluding the wires between the bottom addition gates and the input gates is $\bigO{nmd^3(n^3+m^3)}$.
   Replacing the $nm$ input gates of $C$ by the $\bigO{\ell}$ new input gates $\overline{y}$, $z$, and $1$, and further connecting these gates to the $\bigO{nmd^3(n^3+m^3)}$ bottom addition gates, we obtain a depth-three $f$-oracle circuit over $\F$ of size $\bigO{nmd^3\ell(n^3+m^3)}$ computing 
    \[
        h(\overline{y}, z) \defeq(K_\sigma \mid K_\sigma)(A(\overline{y}, z)) = (1 + \widehat{g}(\overline{y}, z))^t
    \]
    where $t \defeq \abs{\set{i \mid \sigma_i \geq r}} \geq 1$.
    Note that $t \leq \widehat{\sigma}_1 \leq d$.
    
    First, suppose that $\characteristic(\F) = 0$.
    Let $\delta$ be a new indeterminate.
    Then under the map $y_i \mapsto \delta \cdot y_i$ and $z \mapsto \delta$, we have that
    \begin{gather}
    \begin{aligned}\label{eq:det_computation_char_0}
        h(\delta \cdot \overline{y}, \delta) &=  (1 + \widehat{g}(\delta \cdot \overline{y}, \delta))^t \\
                                             &=  (1 + \delta^{\deg(\widehat{g})} \widehat{g}(\overline{y}, 1))^t \\
                                             &= (1 + \delta^{\deg(\widehat{g})} g(\overline{y}))^t \\
                                             &= \sum_{i = 0}^t \binom{t}{i} \delta^{i \cdot \deg(\widehat{g})} g(\overline{y})^i = 1 +  \delta^{\deg(\widehat{g})} g(\overline{y}) + \cdots.
    \end{aligned}
    \end{gather}
    Then $\deg_\delta(h(\delta \cdot \overline{y}, \delta)) \leq t \cdot \deg(\widehat{g}) \leq dr$.
    For each $\alpha\in\F$, we can construct from the depth-three $f$-oracle circuit computing $h(\overline{y},z)$ another $f$-oracle circuit $C_\alpha$ computing $h(\alpha \overline{y}, \alpha)$, whose size is $\bigO{nmd^3\ell(n^3+m^3)}$ and top gate is an addition gate.
    Also, as $\abs{\F} \geq c_0 d^3(n^3+m^3)$, where $c_0>0$ is a large enough constant, we may assume $\abs{\F} \geq dr+1$.
    Therefore, by \Cref{lem:homogenization}, we can compute $\coeff_{\delta^{\deg(\widehat{g})}}(h(\delta\cdot \overline{y},\delta))=g(\overline{y})$ using a depth-three $f$-oracle circuit of size 
    \[
    \bigO{\deg_\delta(h(\delta \cdot \overline{y}, \delta))\cdot nmd^3\ell(n^3+m^3)}=\bigO{nmd^4\ell r(n^3+m^3)}.
    \]

    Now suppose that $\characteristic(\F) = p > 0$.
    The above computation only needs to be modified in the case that $p \mid t$.
    Let $k \in \N$ be the largest natural number such that $t = p^k b$ for some natural number $b$.
    Since $p^k \leq t \leq d$, we have that $k \leq \floor{\log_p(d)}$ as claimed.
    Then by a similar computation to \Cref{eq:det_computation_char_0}, we get that
    \begin{equation}\label{eq:det_computation_char_p} 
        h(\delta \cdot \overline{y}, \delta) = \sum_{i = 0}^t \binom{t}{i} \delta^{i \cdot \deg(\widehat{g})} g(\overline{y})^i = 1 + b \cdot \delta^{\deg(\widehat{g}) p^k} g(\overline{y})^{p^k} + \cdots.
    \end{equation}
    Again, $\deg_\delta(h(\delta \cdot \overline{y}, \delta)) \leq t \cdot \deg(\widehat{g}) \leq dr$.
    For each $\alpha \in \F$, we can construct from the depth-three $f$-oracle circuit computing $h(\overline{y},z)$ another $f$-oracle circuit $C_\alpha$ computing $h(\alpha \overline{y}, \alpha)$, whose size is $\bigO{nmd^3\ell(n^3+m^3)}$ and top gate is an addition gate.
    Also, as $\abs{\F} \geq c_0 d^3(n^3+m^3)$, where $c_0>0$ is a large enough constant, we may assume $\abs{\F} \geq dr+1$.
    Therefore, by \Cref{lem:homogenization}, we have a depth-three $f$-oracle circuit of size $\bigO{nmd^4\ell r(n^3+m^3)}$ computing $\coeff_{\delta^{\deg(\widehat{g})p^k}}(h(\delta\cdot \overline{y},\delta))=b\cdot g(\overline{y})^{p^k}$.
    By dividing the multipliers on the wires connecting to the top gate by $b$, this circuit computing $b\cdot g(\overline{y})^{p^k}$ can be turned into a depth-three $f$-oracle circuit with the same underlying graph that computes $g(\overline{y})^{p^k}$.
\end{pf}

The following corollaries are a debordering of the results~\cite[Corollary 3.9, Corollary 3.10]{AF21} of Andrews and Forbes.
In particular, we can use \Cref{thrm:ABP_det_oracle} to construct a small constant-depth $f$-oracle circuit for the determinant, which partially resolves \Cref{conj:grochow_det} posed by Grochow as stated before.
\begin{cor}\label{cor:det_oracle_for_det}
    Let $f(X) \in I^{\det}_{n, m, r}$ be a nonzero polynomial of degree $d$.
    Let $t=\bigO{r^{1/3}}$ and let $Y$ be a $t \times t$ matrix of indeterminates.
    Assume $\abs{\F} \geq c_0 d^3(n^3+m^3)$, where $c_0>0$ is a large enough constant. 
    Then there is a depth-three $f$-oracle circuit $\Phi$ of size $\bigO{n m d^4 r t^2 (n^3+m^3)} \leq \bigO{n m d^4 r^{5/3} (n^3+m^3)}$ over $\F$ such that 
    \begin{itemize}
        \item if $\characteristic(\F) = 0$ then $\Phi$ computes $\det_t(Y)$, and
        \item if $\characteristic(\F) = p > 0$ then $\Phi$ computes $\det_t(Y)^{p^k}$ for some $k \leq \floor{\log_p(d)}$.
    \end{itemize}
    Furthermore, the top gate of this circuit is an addition gate.
\end{cor}
\begin{pf}
    Mahajan and Vinay~\cite[Theorem 2]{MV97} construct an algebraic branching program on $\bigO{t^3} \leq r$ vertices which computes $\det_t(Y)$.
    Then, the total number of variables in $Y$ is $t^2 = \bigO{r^{2 / 3}}$.
    The result then follows by applying \Cref{thrm:ABP_det_oracle}.
\end{pf}
We can also apply \Cref{thrm:ABP_det_oracle} to other polynomials which are important in the setting of algebraic complexity.
\begin{cor}\label{cor:det_oracle_for_IMM}
    Let $f(X) \in I^{\det}_{n, m, r}$ be a nonzero polynomial of degree $d$.
    Let $W, D$ be positive integers such that $W(D - 1) + 2 \leq r$ and let $Y = Y_1 \sqcup \cdots \sqcup Y_W$ be a set of indeterminates such that each $Y_i$ is a $D \times D$ matrix of indeterminates.
    Let $\IMM_{W, D}$ be the iterated matrix multiplication polynomial, which is the entry in position $(1, 1)$ of the matrix product $Y_1 \cdots Y_W$.
    Assume $\abs{\F} \geq c_0 d^3(n^3+m^3)$, where $c_0>0$ is a large enough constant.
    Then there is a depth-three $f$-oracle circuit $\Phi$ of size $\bigO{nmd^4 r(n^3+m^3) W D^2}$ defined over $\F$ such that 
    \begin{itemize}
        \item if $\characteristic(\F) = 0$ then $\Phi$ computes $\IMM_{W, D}(Y)$, and
        \item if $\characteristic(\F) = p > 0$ then $\Phi$ computes $\IMM_{W, D}(Y)^{p^k}$ for some $k \leq \floor{\log_p(d)}$.
    \end{itemize}
    Furthermore, the top gate of this circuit is an addition gate.
\end{cor}
\begin{pf}
    We have that $\IMM_{W, D}(Y)$ has an algebraic branching program with $W(D-1) + 2 \leq r$ vertices computing it.
    Then, the total number of variables in $Y$ is $WD^2$.
    The result then follows by applying \Cref{thrm:ABP_det_oracle}.
\end{pf} 

\section{Pfaffian Ideals}\label{sec:pfaff}

We will now mimic the determinantal results in \Cref{sec:det} for the case of the ideal $I^{\pfaff}_{2n, 2r}$ of $2r \times 2r$ principal Pfaffians of a $2n \times 2n$ skew-symmetric matrix.
Our application of the straightening law in the skew-symmetric case (\Cref{cor:pfaff-basis}) will be to isolate a canonical bitableau from the expansion of nonzero $f \in I^{\pfaff}_{2n, 2r}$.
In particular, we will show that this can be done via a polynomial-sized depth-three $f$-oracle circuit using the isolation lemma (\Cref{cor:isolation_exponents}) as well as coefficient extraction (\Cref{lem:homogenization}).
This is in contrast to the proof of~\cite{AF21}, where they use a Kronecker-type substitution in order to isolate terms.

Throughout, fix a field $\F$.
Let $n$ be a natural number, and for $1 \leq i < j \leq 2n$, let $x_{i, j}$ be indeterminates.
Throughout this section, let $X = (x_{i, j})$ be a $2n \times 2n$ skew-symmetric matrix of variables so that for all $1 \leq i < j \leq 2n$ we have that $x_{j, i} \defeq -x_{j, i}$ and for all $1 \leq i \leq 2n$ we have that $x_{i, i} \defeq 0$.
To maintain notation with the prior section, we will write $\F[X] = \F[x_{i, j} \mid 1 \leq i < j \leq 2n]$.
Let $1 \leq r \leq n$ and define $I^{\pfaff}_{2n, 2r}$ to be the ideal in $\F[X]$ generated by principal Pfaffians of size $2r$.
As stated in \Cref{defn:principal_bitableau}, we will assume that all partitions, tableaux, and bipfaffians of shape $\sigma$ are such that every row of $\sigma$ has even length.
We will omit this assumption throughout this section.
The proofs below are similar to the ones in \Cref{sec:det} and as such some details may be omitted.

The following is proved in the same manner as \Cref{lem:Eij_left_mult} by studying the case of a single row, making use of \Cref{lem:pfaff_conj}, and then iterating that argument down each row.
\begin{lem}[{\cf~\cite[proof of Lemma 4.1]{AF21}}]\label{lem:Eij_conj}
    Recall the definition of the elementary matrices $E_{i, j}$ from \Cref{defn:elem_mat}.
    Let $\lambda$ be a new indeterminate.
    Let $S$ be a tableau, not necessarily standard, of shape $\sigma$.
    For $0 \leq h \leq h_{i}^{j}(S) - 1$, let $\mathcal{C}^{h}_{i \to j}(S)$ be the set of tableaux of shape $\sigma$ obtained by changing $i$ to $j$ at exactly $h$ rows of $S$ which contain $i$ but not $j$ and reordering those rows to be increasing.
    Then we have that
    \begin{equation}\label{eq:Eij_conj}
        [S](E_{i, j}(\lambda) X E_{i, j}(\lambda)^\top) = \pm \lambda^{h_{i}^{j}(S)} [\Sub_{i \to j}(S)](X) + \sum_{h = 0}^{h_{i}^{j}(S) - 1} \lambda^h \sum_{S' \in \mathcal{C}^{h}_{i \to j}(S)} \pm  [S'](X).
    \end{equation}
\end{lem}

\subsection{Isolating One Term}\label{sec:extract_pfaff_NEW}

In this subsection, in addition to the variables $x_{i,j}$, where $i, j \in [2n]$, we introduce a new set of indeterminates $\Lambda = \set{\lambda_{i, j} | 1 \leq i < j \leq 2n}$.
We have that $\abs{\Lambda} = \binom{n}{2} = \bigO{n^2}$.

Define $M \in \F[\Lambda]^{2n \times 2n}$ by
\begin{align*}
M ={}& E_{1, 2}(\lambda_{1, 2}) \cdots E_{1, 2n}(\lambda_{1, 2n}) E_{2, 3}(\lambda_{2, 3}) \cdots E_{2, 2n}(\lambda_{2, 2n}) \\ 
&\cdots E_{2n - 2, 2n - 1}(\lambda_{2n - 2, 2n - 1}) E_{2n - 2, 2n}(\lambda_{2n - 2, 2n}) E_{2n - 1, 2n}(\lambda_{2n - 1, 2n}).
\end{align*}

We order the elements of $\binom{[2n]}{2} = \set{(i, j) | 1 \leq i < j \leq 2n}$ such that $(i,j) \preceq (i',j')$ if either $i < i'$, or $i = i'$ and $j \leq j'$.
This induces a lexicographic order on $(\Z_{\geq 0})^{\binom{[2n]}{2}}$ which in turn induces a lexicographic term order on the monomials in the variables $\Lambda$ via their degree vectors.
We use $M$ to obtain a polynomial in $\F[X, \Lambda]$ with some terms consisting of anti-canonical bipfaffians.

\begin{lem}\label{lem:pfaffian_leading_terms_NEW}
Let $f(X) \in I^{\det}_{2n, 2r}$ be a nonzero polynomial of degree $d$.
Then $f(MXM^\top)$ can be expressed as a sum
\begin{equation}\label{eq:expansion1_pfaff}
    f(MXM^\top) = \sum_{k \in A} \widetilde{c}_{k} \Lambda^{\overline{e}_k}\cdot [\overline{K}_{\sigma_k}](X) + \sum_{\ell\in A'} \widetilde{c}_{\ell} \Lambda^{\overline{e}_\ell}\cdot [S_\ell](X),
\end{equation}
such that the following hold:
\begin{enumerate}
    \item\label{item:expansion1_i0_pfaff} $A$ is a nonempty finite set and $A'$ is finite set disjoint from $A$.
    \item\label{item:expansion1_i1_pfaff} All terms in \Cref{eq:expansion1_pfaff} have the form $c \Lambda^{\overline{e}}\cdot [S](X)$, where $c\in \F^\times$, $\overline{e}\in \set{0, 1, \dots, d}^{\binom{[2n]}{2}}$, and $S$ has shape $\sigma$ such that $\sigma_1 \geq 2r$ and $\abs{\sigma} \leq 2d$.
    \item\label{item:expansion1_i1.5_pfaff} For every $\ell\in A'$, there exists $k=k(\ell)\in A$ such that $\overline{e}_\ell$ is strictly smaller than $\overline{e}_k$ in the lexicographic order and the shape $\sigma_\ell$ of $[S_\ell]$ equals the shape $\sigma_k$.
    \item\label{item:expansion1_i2_pfaff} The triples $(\overline{e}_k,\sigma_k)$ are distinct, where $k$ ranges over $A$.
\end{enumerate} 
\end{lem}

\begin{pf}
By \Cref{cor:pfaff-basis} and \Cref{prop:pfaff_shape}, we can expand $f(X)$ as a linear combination of standard bipfaffians over $\F$
\[
    f(X) = \sum_{k\in B} c_k \cdot [S_k](X)
\]
where $B$ is nonempty, each $c_k \in \F^\times$, and $[S_k](X)$ is a standard bipfaffian in $I^{\pfaff}_{2n, 2r}$ of shape $\sigma_k$ such that $(\sigma_k)_1 \geq 2r$ and $\abs{\sigma_k} \leq 2d$, with these standard bipfaffians being distinct as $k$ ranges over $B$.
It follows that 
\[
    f(MXM^\top) = \sum_{k\in B} c_k \cdot [S_k](MXM^\top).
\]
For each $k \in B$, we expand the term $c_k \cdot [S_k](MXM^\top)$ by repeatedly applying \Cref{lem:Eij_conj}.
This expansion shows that $c_k \cdot [S_k](MXM^\top)$ equals $\pm c_k \Lambda^{\overline{e}_k}\cdot [\overline{K}_{\sigma_k}](X)$ with $\overline{e}_k=(h_{2n, i, j}(S_k))_{1\leq i < j \leq 2n} \in \set{0,1,\dots,d}^{\binom{[2n]}{2}}$, plus some terms of the form $\pm \widetilde{c}_{\ell}\Lambda^{\overline{e}_\ell}\cdot [S_\ell](X)$, each with a new index $\ell$, where $\widetilde{c}_{\ell}=\pm c_k$, $[S_\ell](X)$ is a (not necessarily standard) bipfaffian of shape $\sigma_\ell = \sigma_k$, and $\overline{e}_\ell$ is strictly smaller than $\overline{e}_k$ in the lexicographic order.
Note that as $\sigma_\ell=\sigma_k$, their first rows both have length at least $2r$, implying that $[S_\ell](X) \in I^{\pfaff}_{2n, 2r}$ by \Cref{prop:pfaff_shape}.
Also note that $\overline{e}_\ell \in \set{0, 1, \dots, d}^{\binom{[2n]}{2}}$ since a substitution $i \mapsto j$ can be performed at most once per row, \ie, at most $(\widehat{\sigma_\ell})_1 \leq d$ times to any tableau of shape $\sigma_\ell$.
The fact that $(\widehat{\sigma_\ell})_1 \leq d$ follows from the fact that all non-empty rows of $\sigma_\ell$ have length $\geq 2$ by \Cref{cor:pfaff-basis}.
We add $\ell$ to $A'$ for each such term and let $k(\ell) = k$ and add this $k$ to $A$.

Therefore, we have the expansion
\[
    f(MXM^\top) = \sum_{k \in A} \pm{c}_{k} \Lambda^{\overline{e}_k}\cdot [\overline{K}_{\sigma_k}](X) + \sum_{\ell\in A'} \widetilde{c}_{\ell} \Lambda^{\overline{e}_\ell}\cdot [S_\ell](X)
\]
satisfying \Crefrange{item:expansion1_i0_pfaff}{item:expansion1_i1.5_pfaff}.

It remains to prove \Cref{item:expansion1_i2_pfaff}.
Recall that as $k$ ranges over $A$ that the $S_k$ are distinct.
Assume to the contrary that there exist distinct $k_1, k_2\in A$ such that $(\overline{e}_{k_1},\sigma_{k_1})=(\overline{e}_{k_2},\sigma_{k_2})$.
Then if $\sigma_{k_1} = \sigma_{k_2}$ we have that $\overline{K}_{\sigma_1} = \overline{K}_{\sigma_2}$.
But then as $\overline{K}_{\sigma_1} = \overline{K}_{\sigma_2}$ and $\overline{e}_{k_1} = \overline{e}_{k_2}$, by \Cref{lem:sub_lex_injective} we have that $S_{k_1} = S_{k_2}$, a contradiction.
\end{pf}

Let $J_{2n} \in \F^{2n \times 2n}$ be the $2n \times 2n$ matrix with $1$'s along the anti-diagonal and $0$'s elsewhere.
In particular, $J_{2n} = J_{2n}^\top$ and so $J_{2n} X J_{2n}$ is still skew-symmetric..
Note that the map $X \mapsto J_{2n} X J_{2n}$ maps the bipfaffian $[\overline{K}_\sigma](X) \mapsto [K_\sigma](X)$ for any partition $\sigma$.
\begin{cor}\label{cor:reformulated_pfaff}
Let $f(X) \in I^{\det}_{2n, 2r}$ be a nonzero polynomial of degree $d$.
Then $f(M J_{2n} X J_{2n} M^\top)$ can be expressed as a sum
\begin{equation}\label{eq:expansion3_pfaff}
    f(M J_{2n} X J_{2n} M^\top) = \sum_{k \in A} \widetilde{c}_{k} \Lambda^{\overline{e}_k}\cdot [K_{\sigma_k}](X) + \sum_{\ell\in A'} \widetilde{c}_{\ell} \Lambda^{\overline{e}_\ell}\cdot [S_\ell](X),
\end{equation}
such that the following hold:
\begin{enumerate}
    \item\label{item:expansion3_i0_pfaff} $A$ is a nonempty finite set and $A'$ is finite set disjoint from $A$.
    \item\label{item:expansion3_i1_pfaff} All terms in \Cref{eq:expansion3_pfaff} have the form $c \Lambda^{\overline{e}}\cdot [S](X)$, where $c\in \F^\times$, $\overline{e}\in \set{0,1,\dots,d}^{\binom{[2n]}{2}}$, and $S$ has shape $\sigma$ such that $\sigma_1 \geq 2r$ and $\abs{\sigma} \leq 2d$.
    \item\label{item:expansion3_i1.5_pfaff} For every $\ell\in A'$, there exists $k=k(\ell)\in A$ such that $\overline{e}_\ell$ is strictly smaller than $\overline{e}_k$ in the lexicographic order and the shape $\sigma_\ell$ of $[S_\ell]$ equals the shape $\sigma_k$.
    \item\label{item:expansion3_i2_pfaff} The triples $(\overline{e}_k,\sigma_k)$ are distinct, where $k$ ranges over $A$.
\end{enumerate} 
\end{cor}

Next, we introduce new variables $Y=\set{y_1,\dots,y_{2n}}$.
Define the diagonal matrix
\[
    D=\mathrm{diag}(y_1,\dots,y_{2n}) \in \F[Y]^{2n \times 2n}.
\]
Note that $D = D^\top$ and that $D X D$ is still skew-symmetric and so taking bipfaffians is still well defined.
\begin{lem}\label{lem:diagonals_pfaff}
Let $[S](X)$ be a bideterminant of degree $d$.
Then 
\[
    [S](D X D)= Y^{\overline{s}} [S](X),
\]
where $\overline{s}=(s_1,\dots,s_{2n})\in \set{0, 1, \dots, 2d}^{2n}$, $s_i$ is the number of times $i$ appears in $S$, \ie, $s_1 e_1+\dots+s_{2n} e_{2n}$ is the multidegree of $[S](X)$ with respect to the grading defined in \Cref{defn:multideg_pfaff}.
\end{lem}
\begin{pf}
    Reduce to the case in which $S$ has only one row, and then use the multilinearity of the determinant and the fact that the Pfaffian is the square-root of the determinant.
\end{pf}

We will often simply say that $[S](X)$ in \Cref{lem:diagonals_pfaff} has multidegree $\overline{s}$ (rather than writing it as $s_1 e_1 + \dots + s_n e_n$).

Next, we introduce another variable $v$ and modify $D$:
\[
    \widetilde{D} = \mathrm{diag}(y_1 v, y_2 v^2\dots,y_{2n} v^{2n} ) \in \F[Y,v]^{2n\times 2n}.
\]
Note that $\widetilde{D} = \widetilde{D}^\top$ and that $\widetilde{D} X \widetilde{D}$ is still skew-symmetric and so taking bipfaffians is still well defined.
The proof of the following lemma is similar to the one of \Cref{lem:multiply-by-diag} and is thus omitted.
\begin{lem}\label{lem:multiply-by-diag_pfaff}
Let $[S](X)$ be a bipfaffian of degree $d$.
Then 
\[
    [S](\widetilde{D} X \widetilde{D}) = Y^{\overline{s}}v^{\abs{S}} \cdot [S](X),
\]
where $\overline{s}$ is the multidegree of $[S](X)$.
\end{lem}

\begin{lem}\label{lem:key_pfaff}
Let $f(X) \in I^{\pfaff}_{2n, 2r}$ be a nonzero polynomial of degree $d$.
Let $g = f(M J_{2n} \widetilde{D} X \widetilde{D} J_{2n} M^\top) \in \F[X, \Lambda, Y, v]$.
View $g$ as a univariate polynomial in $v$ with coefficients in $\F[X, \Lambda, Y]$, and write $g=\sum_{i} \coeff_{v^i}(g) v^i$, where $\coeff_{v^i}(g)\in \F[X, \Lambda ,Y]$ denotes the coefficient of $v^i$ in $g$.
Choose the smallest integer $d_{\min} = \bigO{dn}$ such that  $\coeff_{v^{d_{\min}}}(g)\neq 0$.
Then $d_{\min} = \abs{K_\sigma}$ for some shape $\sigma$ with $\abs{\sigma}\leq 2d$ and we may write $\coeff_{v^{d_{\min}}}(g)$ as a finite sum
\[
    \coeff_{v^{d_{\min}}}(g) = \sum_{k\in I} c_k \Lambda^{\overline{e}_k} Y^{\overline{s}_k} \cdot [K_{\sigma_k}](X), 
\]
such that the following hold:
\begin{enumerate}
    \item\label{item:required1_pfaff} For each $k\in I$, $c_k\in\F^\times$,  $(\sigma_k)_1\geq 2r$, and $\abs{\sigma_k} \leq 2d$.
    \item\label{item:required2_pfaff} The tuples $(\overline{e}_k, \overline{s}_k)$ are distinct, where $k$ ranges over $I$.
\end{enumerate}
\end{lem}

\begin{pf}
Consider the expansion
\[
        f(M J_{2n} X J_{2n} M^\top) = \sum_{k \in A} \widetilde{c}_{k} \Lambda^{\overline{e}_k}\cdot [K_{\sigma_k}](X) + \sum_{\ell\in A'} \widetilde{c}_{\ell} \Lambda^{\overline{e}_\ell}\cdot [S_\ell](X),
\]
given by \Cref{cor:reformulated_pfaff}.
Then we have that
\begin{equation}\label{eq:expansion-g_pfaff}
\begin{aligned}
    g &= \sum_{k \in A} \widetilde{c}_{k} \Lambda^{\overline{e}_k}\cdot [K_{\sigma_k}](\widetilde{D} X \widetilde{D}) + \sum_{\ell\in A'} \widetilde{c}_{\ell} \Lambda^{\overline{e}_\ell}\cdot [S_\ell](\widetilde{D} X \widetilde{D}) \\
      &= \sum_{k \in A} \widetilde{c}_{k} \Lambda^{\overline{e}_k} Y^{\overline{s}_k} v^{\abs{K_{\sigma_k}}} \cdot [K_{\sigma_k}](X) + \sum_{\ell\in A'} \widetilde{c}_{\ell} \Lambda^{\overline{e}_\ell} Y^{\overline{s}_\ell} v^{\abs{S_\ell}} \cdot [S_\ell](X),
\end{aligned}
\end{equation}
where for $k\in A$, $\overline{s}_k$ is the multidegree of $[K_{\sigma_k}](X)$, and for $\ell \in B$, $\overline{s}_\ell$ is the multidegree of $[S_\ell](X)$.
Here, the second equality holds by \Cref{lem:multiply-by-diag_pfaff}.

Choose $k_0 \in A$ so that $\abs{K_{\sigma_{k_0}}}$ is minimized, and subject to this, $\overline{e}_{k_0}$ is maximized in the lexicographic order.
Let $d_0 = \abs{K_{\sigma_{k_0}}}$.
We will show that $\coeff_{v^{d_0}}(g)\neq 0$.

For any $k\in A$, consider the term $\widetilde{c}_{k} \Lambda^{\overline{e}_k} Y^{\overline{s}_k} v^{\abs{K_{\sigma_k}}} \cdot [K_{\sigma_k}](X)$ in \Cref{eq:expansion-g_pfaff}.
It is homogeneous of degree $\abs{K_{\sigma_k}} \geq \abs{K_{\sigma_{k_0}}} = d_0$ with respect to $v$ by the minimality of $\abs{K_{\sigma_{k_0}}}$, and contributes to $\coeff_{v^{d_0}}(g)$ if and only if $\abs{K_{\sigma_k}} = \abs{K_{\sigma_{k_0}}}$.
Even if it can potentially contribute to $\coeff_{v^{d_0}}(g)$, its (multi)degree $(\overline{e}_k, \overline{s}_k)$ in the variables $\Lambda$ and $Y$ is different from that of the term indexed by $k_0$ unless $k = k_0$.
To see this, suppose $k \neq k_0$.
Then $(\overline{e}_k, \sigma_k) \neq (\overline{e}_{k_0} ,\sigma_{k_0})$.
If $\sigma_k = \sigma_{k_0}$, then $\overline{e}_k \neq \overline{e}_{k_0}$.
On the other hand, if $\sigma_k \neq \sigma_{k_0}$, then $\overline{s}_k \neq \overline{s}_{k_0}$ by \Cref{lem:canonical_degs_bifpaffian}.
Thus, we always have that
\[
    (\overline{e}_k, \overline{s}_k) \neq (\overline{e}_{k_0},\overline{s}_{k_0}).
\]

Now consider $\ell\in B$ and the term $\mathfrak{m}_\ell \defeq \widehat{c}_{\ell} \Lambda^{\overline{e}_\ell} Y^{\overline{s}_\ell} v^{\abs{S_\ell}} \cdot [S_\ell](X)$ in \Cref{eq:expansion-g_pfaff}.
The degree of $\mathfrak{m}_\ell$ in $v$ is $\abs{S_\ell}$.
Let $k = k(\ell)$ as in \Cref{cor:reformulated_pfaff}.
Then by \Cref{cor:reformulated_pfaff}~\eqref{item:expansion1_i1.5_pfaff}, $S_\ell$ has shape $\sigma_\ell = \sigma_k$.
Note that among all bipfaffians $[S]$ of shape $\sigma$, the quantity $\abs{S}$ is minimized when, and only when, $[S] = [K_\sigma]$.
Thus, we have 
\[
    \abs{S_\ell} \geq \abs{K_{\sigma_k}} \geq \abs{K_{\sigma_{k_0}}} = d_0.
\]
and $\deg_v(\mathfrak{m}_\ell)= \abs{S_\ell} = d_0$ if and only if $[S_\ell] = [K_{\sigma_k}]$ and $\abs{K_{\sigma_{k}}} = \abs{K_{\sigma_{k_0}}}$.
Even though this can happen, when it happens we have that $\overline{e}_k$ is smaller than or equal to $\overline{e}_{k_0}$ in the lexicographic order by the choice of $k_0$, noting that $\abs{K_{\sigma_{k}}} = \abs{K_{\sigma_{k_0}}}$.
And by \Cref{cor:reformulated_pfaff}~\eqref{item:expansion1_i1.5_pfaff}, $\overline{e}_\ell$ is strictly smaller than $\overline{e}_k$ in the lexicographic order.
Overall, either $\deg_v(\mathfrak{m}_\ell) > d_0$, or $(\overline{e}_\ell,\overline{s}_\ell) \neq (\overline{e}_{k_0}, \overline{s}_{k_0})$.

By the above analysis, the term $\widehat{c}_{k_0} \Lambda^{\overline{e}_{k_0}} Y^{\overline{s}_{k_0}} v^{\abs{K_{\sigma_{k_0}}}} \cdot [K_{\sigma_{k_0}}](X)$ is not canceled by other terms contributing to $\coeff_{v^{d_0}}(g)$ due to the uniqueness of its (multi)degree $(\overline{e}_{k_0}, \overline{s}_{k_0})$ in $\Lambda$ and $Y$.
Thus $\coeff_{v^{d_0}}(g)\neq 0$.
The above analysis also shows that $\coeff_{v^{i}}(g)=0$ for $i < d_0$.
Thus, $d_0 = d_{\min}$.
Note that $d_{\min} = \abs{K_{\sigma_{k_0}}} \leq \bigO{dn}$ as each of the at most $2d$ boxes in $K_{\sigma_{k_0}}$ have entry at most $2n$.

Moreover, the above analysis shows that any term that contributes to $\coeff_{v^{d_0}}(g)$ contains a bipfaffian of the form $[K_\sigma](X)$.
We then merge the terms contributing to $\coeff_{v^{d_0}}(g)$ with the same (multi)degree in $\Lambda$ and $Y$.
When we merge two terms $\widehat{c}_{k} \Lambda^{\overline{e}_k} Y^{\overline{s}_k} v^{d_0} \cdot [K_{\sigma_k}](X)$ and $\widehat{c}_{k'} \Lambda^{\overline{e}_{k'}} Y^{\overline{s}_{k'}} v^{d_0} \cdot [K_{\sigma_{k'}}](X)$ with $(\overline{e}_k,\overline{s}_k) = (\overline{e}_{k'}, \overline{s}_{k'})$, we always have $[K_{\sigma_{k}}] = [K_{\sigma_{k'}}]$.
This follows from \Cref{lem:canonical_degs_bifpaffian} and the fact that $\overline{s}_{k}$ is the multidegree of $[K_{\sigma_{k}}](X)$ and $\overline{s}_{k'}$ is the multidegree of $[K_{\sigma_{k'}}]$.
Thus, merging the two terms yields a scalar multiple of both.

After merging/canceling terms, we can write $\coeff_{v^{d_{\min}}}(g) = \coeff_{v^{d_0}}(g)$ in the form of \Cref{eq:required} such that \Cref{item:required2} holds.
Furthermore, \Cref{item:required1} holds as well by \Cref{cor:reformulated_pfaff}~\eqref{item:expansion1_i1_pfaff} and \Cref{prop:pfaff_shape}.
\end{pf}

\Cref{lem:key_pfaff} helps us separate a collection of terms solely containing canonical bipfaffians.
The next lemma further applies the isolation lemma to single out one term from this collection.

\begin{lem}\label{lem:univariate_pfaff}
Let $f(X) \in I^{\pfaff}_{2n, 2r}$ be a nonzero polynomial of degree $d$.
Let $g = f(M J_{2n} \widetilde{D} X \widetilde{D} J_{2n} M^\top) \in \F[X, \Lambda, Y, v]$.
Then there exist integers $z_t = \bigO{dn^2}$ for each variable $t\in \Lambda \sqcup Y$, and an integer $z_v = \bigO{d^2 n^2}$ such that for the two variable substitution maps 
\[
    \phi\colon t \mapsto w^{z_t} \quad\text{ for } t\in \Lambda \sqcup Y
\]
and
\[
    \psi\colon v \mapsto w^{z_v},
\]
we have that $h \defeq (\psi\circ\phi)(g)\in\F[X,w]$ has $\deg_w(h) = \bigO{d^3 n^3}$ such that $\coeff_{w^i}(h) = c\cdot [K_\sigma]$ for some integer $i \leq \deg_w(h)$, where $c\in\F^\times$, $\sigma_1\geq 2r$ and $\abs{\sigma}\leq 2d$.
\end{lem}
\begin{pf}
Consider the finite sum from \Cref{lem:key_pfaff}:
\[
    \coeff_{v^{d_{\min}}}(g) = \sum_{k\in I} c_k \Lambda^{\overline{e}_k} Y^{\overline{s}_k} \cdot [K_{\sigma_k}](X)
\]
The coordinates of $\overline{e}_k$ and $\overline{s}_k$ are in $\set{0,1,\dots,d}$ and $\set{0, 1, \dots, 2d}$ respectively for each $k \in I$ by \Cref{cor:reformulated_pfaff}~\eqref{item:expansion1_i1_pfaff} and \Cref{lem:multiply-by-diag_pfaff}.
Choose a sufficiently large $L = \bigTheta(dn^2)$, and pick integers $z_t$ independently and uniformly at random from $\set{0, 1, \dots, L}$, where $t\in \Lambda \sqcup Y$.
Then by \Cref{cor:isolation_exponents}, with high probability, there exists an integer $d'$ such that the coefficient of the monomial $v^{d_{\min}}w^{d'}$ in $\phi(g) \in \F[X,v,w]=\F[X][v,w]$ has the form $c\cdot [K_\sigma](X) \in \F[X]$, where $c\in\F^\times$, $\sigma_1 \geq 2r$, and $\abs{\sigma} \leq 2d$.
Fix the integers $z_t$ such that this occurs.

By \Cref{lem:multiply-by-diag_pfaff}, the degree of $\phi(g)$ in $v$ is at most $\bigO{dn}$ as each of the at most $2d$ boxes has entry at most $2n$.
The degree of $\phi(g)$ in $w$ is bounded by $\bigO{dL} = \bigO{d^2 n^2}$.
Choose $z_v=\deg_w(\phi(g))+ 1 = \bigO{d^2 n^2}$.
Then the map $\psi\colon v\mapsto w^{z_v}$ sends the monomials of $\phi(g)$ bijectively to that of $(\psi\circ\phi)(g)=h$, preserving coefficients.
Thus, there exists an integer $i$ such that $\coeff_{w^i}(h) = c\cdot [K_\sigma](X)$.
Finally, we have that
\[
    \deg_w(h) \leq z_v \cdot \bigO{dn} + d' \leq \bigO{d^2 n^2} \cdot \bigO{dn} + \bigO{d^2 n^2} = \bigO{d^3 n^3}.
\]
\end{pf}

We are now ready to prove the main theorem in this subsection.

\begin{thrm}\label{thrm:bipfaffian_canonical_extraction}
    Let $f(X) \in I^{\pfaff}_{2n, 2r}$ be a nonzero polynomial of degree $d$.
    Assume $\abs{\F} \geq c_0 d^3 n^3$, where $c_0>0$ is a large enough constant.
    Then there exists a depth-three $f$-oracle circuit of size $\bigO{d^3 n^7} = \poly(n, d)$ computing $[K_\sigma](X)$, where $\sigma_1\geq 2r$ and $\abs{\sigma} \leq 2d$.
    Furthermore, the top gate of this circuit is an addition gate, and the bottom layer consists of $\bigO{d^3 n^5}$ addition gates.
    The total number of gates and wires excluding the wires between the bottom addition gates and the input gates is $\bigO{d^3 n^5}$.
\end{thrm}
\begin{pf}
Let $h$ be as in \Cref{lem:univariate_pfaff}.
Let $\alpha \in \F$ be arbitrary.
By construction, $h(X,\alpha)$ equals $f(M_\alpha X M_\alpha^{\top})$, where $M_\alpha$ is obtained from $M J_{2n} \widetilde{D}$ by substituting powers of $\alpha$ for the variables $\Lambda$, $Y$, and $v$.
Specifically, if under $\psi \circ \phi$ a variable is mapped to $w^s$ for some integer $s \geq 0$, then we substitute $\alpha^s$ for that variable here.
As the map $X\mapsto M_\alpha X M_\alpha^{\top}$ is a linear map, we can construct a depth-two $f$-oracle circuit $C_\alpha$ computing $h(X,\alpha) = f(M_\alpha X N_\alpha)$ as follows:
Use $\bigO{n^2}$ addition gates and $\bigO{n^4}$ wires at the bottom layer computing the $4n^2$ entries of $M_\alpha X M_\alpha^{\top}$.
The top gate is an $f$-gate connecting to the $n^2$ addition gates via $n^2$ wires.
Thus, the size of $C_\alpha$ is $\bigO{n^4}$.

Let $d_w=\deg_w(h)$.
By \Cref{lem:univariate_pfaff}, we have that $d_w = \bigO{d^3 n^3}$ and that $\coeff_{w^i}(h) = c\cdot [K_\sigma]$ for some integer $i \leq \deg_w(h)$, where $c\in\F^\times$, $\sigma_1\geq 2r$ and $\abs{\sigma}\leq 2d$.
As $\abs{\F} \geq c_0 d^3n^3$, where $c_0>0$ is a large enough constant, we may assume $\abs{\F} \geq d_w+1$.
Then by \Cref{lem:homogenization}, we have a depth-three $f$-oracle circuit $C$ of size $\bigO{d_w n^4} = \bigO{d^3 n^7}$ computing $\coeff_{w^i}(h)=c\cdot (K_\sigma\mid K_\sigma)$.
The top gate of $C$ is an addition gate, which connects to $d_w+1 = \bigO{d^3 n^3}$ $f$-gates on the middle layer.
Then, these $f$-gates connect to $(d_w+1)n^2 = \bigO{d^3 n^5}$ addition gates on the bottom layer.

The above circuit $C$ computes $c\cdot [K_\sigma]$.
But by dividing the constants on the wires connecting the top addition gate by $c$, we can transform $C$ into an $f$-oracle circuit with the same underlying graph that computes $[K_\sigma](X)$.
This proves the theorem.
\end{pf}

\subsection{Expressing Algebraic Branching Programs as Pfaffians of a Skew-Symmetric Matrix}

The remainder of the proof follows~\cite{AF21}, which we include here for completeness.
One key idea is exploiting the close connection between two models: algebraic branching programs and determinants.
In particular, we will need the following lemma:

\begin{lem}[{\cite[Lemma 4.3]{AF21}}]\label{lem:skew_symmetrization}
  Let $A$ be an $n\times n$ matrix.
  Then there is a $2n \times 2n$ skew-symmetric matrix $M$ such that for all $k \in [n]$, we have that $\pfaff_{2k}(M_{[2k], [2k]}) = \pm \det_k(A_{[k], [k]})$.
\end{lem}

\subsection{Computing the Pfaffian of a Skew-Symmetric Matrix}

We will use \Cref{thrm:bipfaffian_canonical_extraction} to obtain a constant-depth algebraic circuit of size $\poly(n,\deg(f))$ with $f$-oracle gates that computes the Pfaffian of a skew-symmetric symbolic matrix, thereby resolving the analogue of \Cref{question:main} for the Pfaffian case. We start by proving the following more general theorem.
\begin{thrm}\label{thrm:ABP_pfaff_oracle}
    Let $f(X) \in I^{\pfaff}_{2n, 2r}$ be a nonzero polynomial of degree $d$.
    Let $g(y_1, \ldots, y_\ell) \in \F[\overline{y}]$ be a polynomial computable by an algebraic branching program with at most $r$ vertices.
    Assume $\abs{\F} \geq c_0 d^3 n^3$, where $c_0>0$ is a large enough constant.
    Then there is a depth-three $f$-oracle circuit $\Phi$ of size $\bigO{d^4 \ell n^5 r}$ defined over $\F$ such that 
    \begin{itemize}
        \item if $\characteristic(\F) = 0$ then $\Phi$ computes $g(\overline{y})$, and
        \item if $\characteristic(\F) = p > 0$ then $\Phi$ computes $g(\overline{y})^{p^k}$ for some $k \leq \floor{\log_p(d)}$.
    \end{itemize}
    Furthermore, the top layer of this circuit consists of an addition gate.
\end{thrm}
\begin{pf}
    By \Cref{lem:homog_ABP}, there is a homogeneous polynomial $\widehat{g}(\overline{y}, z) \in \F[\overline{y}, z]$ such that $\widehat{g}(\overline{y}, 1) = g(\overline{y})$ and $\widehat{g}(\overline{y}, z)$ can be computed by an algebraic branching program over $\F$ with at most $r$ vertices.
    We may add isolated vertices such that the algebraic branching program has exactly $r$ vertices.
    Then by \Cref{lem:ABP_to_det}, there is a matrix $A(\overline{y}, z) \in \F[\overline{y}, z]^{r \times r}$ such that
    \begin{itemize}
        \item $\det_r(A(\overline{y}, z)) = 1 + \widehat{g}(\overline{y}, z)$,
        \item $\det_i(A(\overline{y}, z)_{[i], [i]}) = 1$ for all $1 \leq i \leq r - 1$,
    \end{itemize}
    and all entries of $A(\overline{y}, z)$ have degree at most $1$ in $\overline{y}$ and $z$.  
    Extend $A(\overline{y}, z)$ to an $n\times n$ matrix by adding $1$'s to the main diagonal and $0$'s elsewhere.
     Then, by \Cref{lem:skew_symmetrization}, there exists a $2n \times 2n$ skew-symmetric matrix $M$ such that $\pfaff_{2k}(M_{[2k], [2k]}) = \pm \det_k(A_{[k], [k]})$ for all $1 \leq k \leq n$.
    
    By \Cref{thrm:bipfaffian_canonical_extraction}, there exists a depth-three $f$-oracle circuit $C$ over $\F$ of size $\bigO{d^3 n^7}$ computing $[K_\sigma](X)$ such that $\sigma_1 \geq 2r$ and $\abs{\sigma} \leq 2d$.
    Furthermore, the top gate of $C$ is an addition gate, the bottom layer of $C$ consists of $\bigO{d^3 n^5}$ addition gates, and the total number of gates and wires excluding the wires between the bottom addition gates and the input gates is $\bigO{d^3 n^5}$.
    Replacing the $n^2$ input gates of $C$ by the $\bigO{\ell}$ new input gates $\overline{y}$, $z$, and $1$, and further connecting these gates to the $\bigO{d^3 n^5}$ bottom addition gates, we obtain a depth-three $f$-oracle circuit over $\F$ of size $\bigO{d^3 \ell n^5}$ computing 
    \begin{align*}
        h(\overline{y}, z) &\defeq [K_\sigma](A(\overline{y}, z)) \\
                           &= \prod_{i = 1}^{\widehat{\sigma}_1} \pfaff_{\sigma_i}(A(\overline{y}, z)_{[\sigma_i], [\sigma_i]}) \\
                           &= \pm \prod_{i = 1}^{\widehat{\sigma}_1} \det_{\sigma_i / 2}(A(\overline{y}, z)_{[\sigma_i / 2], [\sigma_i / 2]}) \\
                           &= \pm \prod_{i;\ \sigma_i \geq 2r} \det_{\sigma_i / 2} (A(\overline{y}, z)_{[\sigma_i / 2], [\sigma_i / 2]}) \cdot \prod_{i;\ \sigma_i < 2r} \det_{\sigma_i / 2}(A(\overline{y}, z)_{[\sigma_i / 2], [\sigma_i / 2]}) \\
                           &= \pm (1 + \widehat{g}(\overline{y}, z))^t
    \end{align*}
    where $t \defeq \abs{\set{i \mid \sigma_i \geq 2r}} \geq 1$.
    Note that $t \leq \widehat{\sigma}_1 \leq d$.
    
    First, suppose that $\characteristic(\F) = 0$.
    Let $\delta$ be a new indeterminate.
    Then under the map $y_i \mapsto \delta \cdot y_i$ and $z \mapsto \delta$, we have that
    \begin{gather}
    \begin{aligned}\label{eq:pfaff_computation_char_0}
        h(\delta \cdot \overline{y}, \delta) &= \pm (1 + \widehat{g}(\delta \cdot \overline{y}, \delta))^t \\
                                             &= \pm (1 + \delta^{\deg(\widehat{g})} \widehat{g}(\overline{y}, 1))^t \\
                                             &= \pm (1 + \delta^{\deg(\widehat{g})} g(\overline{y}))^t \\
                                             &= \pm \sum_{i = 0}^t \binom{t}{i} \delta^{i \cdot \deg(\widehat{g})} g(\overline{y})^i = \pm \pqty{1 + \delta^{\deg(\widehat{g})} t \cdot g(\overline{y}) + \cdots}.
    \end{aligned}
    \end{gather}
    Then $\deg_\delta(h(\delta \cdot \overline{y}, \delta)) \leq t \cdot \deg(\widehat{g}) \leq dr$.
    For each $\alpha\in\F$, we can construct from the depth-three $f$-oracle circuit computing $h(\overline{y},z)$ another $f$-oracle circuit $C_\alpha$ computing $h(\alpha \overline{y}, \alpha)$, whose size is $\bigO{d^3 \ell n^5}$ and top gate is an addition gate.
    Also, as $\abs{\F} \geq c_0 d^3 n^3$, where $c_0>0$ is a large enough constant, we may assume $\abs{\F} \geq dr+1$.
    Therefore, by \Cref{lem:homogenization}, we can compute $\coeff_{\delta^{\deg(\widehat{g})}}(h(\delta\cdot \overline{y},\delta))= \pm t \cdot g(\overline{y})$ using a depth-three $f$-oracle circuit of size 
    \[
        \bigO{\deg_\delta(h(\delta \cdot \overline{y}, \delta))\cdot d^3 \ell n^5} = \bigO{d^4 \ell n^5 r}.
    \]
    Dividing the constants on the incoming wires to the top addition gate from \Cref{lem:homogenization} by $\pm t$, we get a circuit using the same underlying graph computing $g(\overline{y})$.

    Now suppose that $\characteristic(\F) = p > 0$.
    The above computation only needs to be modified in the case that $p \mid t$.
    Let $k \in \N$ be the largest natural number such that $t = p^k b$ for some natural number $b$.
    Since $p^k \leq t \leq d$, we have that $k \leq \floor{\log_p(d)}$ as claimed.
    Then by a similar computation to \Cref{eq:pfaff_computation_char_0}, we get that
    \begin{equation}\label{eq:pfaff_computation_char_p}
        h(\delta \cdot \overline{y}, \delta) = \sum_{i = 0}^t \binom{t}{i} \delta^{i \cdot \deg(\widehat{g})} g(\overline{y})^i = \pm(1 + b \cdot \delta^{\deg(\widehat{g}) p^k} g(\overline{y})^{p^k} + \cdots).
    \end{equation}
    Again, $\deg_\delta(h(\delta \cdot \overline{y}, \delta)) \leq t \cdot \deg(\widehat{g}) \leq dr$.
    For each $\alpha\in\F$, we can construct from the depth-three $f$-oracle circuit computing $h(\overline{y},z)$ another $f$-oracle circuit $C_\alpha$ computing $h(\alpha \overline{y}, \alpha)$, whose size is $\bigO{d^3 \ell n^5}$ and top gate is an addition gate.
    Also, as $\abs{\F} \geq c_0 d^3 n^3$, where $c_0>0$ is a large enough constant, we may assume $\abs{\F} \geq dr+1$.
    By \Cref{lem:homogenization}, we have a depth-three $f$-oracle circuit of size $\bigO{d^4 \ell n^5 r}$ computing $\coeff_{\delta^{\deg(\widehat{g})p^k}}(h(\delta\cdot \overline{y},\delta))= \pm \cdot b \cdot g(\overline{y})^{p^k}$.
    By dividing the constants on the wires connecting to the top gate by $\pm b$, this circuit computing $b\cdot g(\overline{y})^{p^k}$ can be turned into a depth-three $f$-oracle circuit with the same underlying graph that computes $g(\overline{y})^{p^k}$.
\end{pf}

The following corollary is a debordering of the result~\cite[Corollary 4.5]{AF21} of Andrews and Forbes.
\begin{cor}\label{cor:pfaff_oracle_for_pfaff}
    Let $f(X) \in I^{\pfaff}_{2n, 2r}$ be a nonzero polynomial of degree $d$.
    Let $t \leq \bigO{\sqrt[3]{r}}$ and let $Y$ be a $t \times t$ matrix of indeterminates.
    Assume $\abs{\F} \geq c_0 d^3 n^3$, where $c_0>0$ is a large enough constant.
    Then there is a depth-three $f$-oracle circuit $\Phi$ of size $\bigO{d^4 t^2 n^5 r} \leq \bigO{d^4 n^5 r^{5 / 3}}$ defined over $\F$ such that
    \begin{itemize}
        \item if $\characteristic(\F) = 0$ then $\Phi$ computes $\pfaff_t(Y)$, and
        \item if $\characteristic(\F) = p > 0$ then $\Phi$ computes $\pfaff_t(Y)^{p^k}$ for some $k \leq \floor{\log_p(d)}$.
    \end{itemize}
    Furthermore, the top layer of this circuit consists of an addition gate.
\end{cor}
\begin{pf}
    Mahajan, Subramany, and Vinay~\cite[Theorem 12]{MSV04} construct an algebraic branching program on $\bigO{t^3} \leq r$ vertices which computes $\pfaff_{t}(Y)$.
    Then, the total number of variables is $t^2 \leq \bigO{r^{2 / 3}}$.
    The result then follows by applying \Cref{thrm:ABP_pfaff_oracle}.
\end{pf}

\section{Conclusions and Open Questions}

In this work, we showed that for any nonzero $f \in I^{\det}_{n,m,r}$ of polynomial degree, the $t\times t$ determinant with $t = \bigTheta{r^{1/3}}$ can be exactly computed by a depth-three, polynomial-size $f$-oracle circuit.
We also established an analogous result for Pfaffian ideals.
These results extend and deborder the work of Andrews and Forbes~\cite{AF21}.
Our results can be viewed as providing non-principal-ideal analogs of the classical closure results for principal ideals, as factoring results for $g \mid f$ can be viewed as closure results for the principal ideal generated by $g$.
However, our results currently only hold for certain non-principal examples.
% Reworded the end of this, maybe some more elaboration as to which results have been extended would be good.

We conclude with several open questions and directions:

\begin{enumerate}
\item In our main theorem \Cref{thrm:main}, the circuit size depends polynomially on the degree $\deg(f)$ of $f$.
Grochow conjectured that this dependence on $\deg(f)$ can be completely removed (see \Cref{conj:grochow_det}).
It would be interesting to see whether the dependence can at least be improved to subpolynomial.
Conversely, if one doubts the conjecture, it would be valuable to identify candidate families of polynomials in determinantal ideals that might serve as counterexamples.

\item Our results, like those of~\cite{AF21}, are proved only over fields of characteristic zero or sufficiently large positive characteristic 
The difficulty in small characteristic is that in \Cref{thrm:ABP_det_oracle}, the $f$-oracle circuit $\Phi$ does not compute $g(\overline{y})$ directly, but only a power of $g(\overline{y})$.
Obtaining a polynomial-sized circuit that computes the $p$-th root of a given circuit over a field of characteristic $p$ is currently feasible only when the number of variables is small~\cite{And20}.
This obstacle is closely related to the fact that the celebrated superpolynomial lower bound for constant-depth algebraic circuits in~\cite{LST25} is not known to imply subexponential-time deterministic PIT algorithms in small positive characteristic, even though the lower bound itself and some of its proof-complexity applications do extend to that setting~\cite{For24,BLRS25}.
Determining whether our results, and those of~\cite{AF21}, can be extended to small positive characteristic therefore remains an important open problem.

\item It is natural to ask whether similar results on the complexity of ideals can be obtained in other settings.
For instance, we expect analogous results for symmetric determinantal ideals (where the matrix variables satisfy $x_{i,j} = x_{j,i}$), though we leave this for future work.
As mentioned in the introduction, determinantal and Pfaffian ideals can all be studied within a unified framework known as Standard Monomial Theory~\cite{LR08,Ses16}.
The key observation is that the corresponding determinantal varieties can be realized as affine open subsets of certain Schubert varieties.
With this connection in place, one can develop the theory---including standard monomial bases and straightening laws---on these Schubert varieties and then restrict to the affine open subsets to obtain the desired statements about the determinantal ideals.
As noted in~\cite{Mus03}, this approach also applies to other affine varieties such as ladder determinantal ideals, varieties of complexes, and quiver varieties, potentially yielding further cases in which one can establish complexity results for ideals.
\end{enumerate}

\section*{Acknowledgments}

We thank David Anderson, Robert Andrews, and Srikanth Srinivasan for helpful discussions. 
We also thank the anonymous reviewers of ITCS 2026 for their careful reading and insightful comments.

\printbibliography

@inproceedings{AF21,
  title = {Ideals, Determinants, and Straightening: Proving and Using Lower Bounds for Polynomial Ideals},
  author = {Andrews, Robert and Forbes, Michael A.},
  booktitle = {Proceedings of the 54th Annual ACM Symposium on Theory of Computing},
  pages = {389--402},
  year = {2022},
  doi = {10.1145/3519935.3520025},
}

@inproceedings{KS01,
  title = {Randomness Efficient Identity Testing of Multivariate Polynomials},
  author = {Klivans, Adam R. and Spielman, Daniel},
  booktitle = {Proceedings of the 33rd Annual ACM Symposium on Theory of Computing},
  pages = {216--223},
  year = {2001},
  doi = {10.1145/380752.380801},
}

@misc{Sap21,
  author = {Ramprasad Saptharishi},
  title = {A Survey of Lower Bounds in Arithmetic Circuit Complexity},
  month = {7},
  year = {2021},
  url = {https://github.com/dasarpmar/lowerbounds-survey/releases},
  note = {Version 9.0.3},
}

@inproceedings{Val79,
  title = {Completeness Classes in Algebra},
  author = {Valiant, Leslie G.},
  booktitle = {Proceedings of the 11th Annual ACM Symposium on Theory of Computing},
  pages = {249--261},
  year = {1979},
  doi = {10.1145/800135.804419},
}

@techreport{MV97,
  author = {Mahajan, Meena and Vinay, V.},
  title = {Determinant: Combinatorics, Algorithms, and Complexity},
  year = {1997},
  number = {5},
  publisher = {Chicago Journal of Theoretical Computer Science},
  institution = {Institute of Mathematical Sciences India},
}

@article{MSV04,
  title = {The Combinatorial Approach Yields an NC Algorithm for Computing Pfaffians},
  author = {Mahajan, Meena and Subramanya, P.R. and Vinay, V.},
  journal = {Discrete Applied Mathematics},
  volume = {143},
  number = {1-3},
  pages = {1--16},
  year = {2004},
  publisher = {Elsevier},
  doi = {10.1016/j.dam.2003.12.001},
}

@article{Grochow20,
  author = {Joshua A. Grochow},
  title = {Complexity in Ideals of Polynomials: Questions on Algebraic Complexity of Circuits and Proofs},
  journal = {Bulletin of the European Association for Theoretical Computer Science},
  volume = {131},
  year = {2020}, 
  url = {http://smtp.eatcs.org/index.php/beatcs/article/view/620},
}

@article{Str73,
  author = {Strassen, Volker},
  journal = {Journal für die reine und angewandte Mathematik},
  pages = {184-202},
  title = {Vermeidung von Divisionen},
  url = {http://eudml.org/doc/151394},
  volume = {264},
  year = {1973},
}

@article{DRS74,
  author = {Doubilet, Peter and Rota, Gian-Carlo and Stein, Joel},
  title = {On the Foundations of Combinatorial Theory: IX Combinatorial Methods in Invariant Theory},
  journal = {Studies in Applied Mathematics},
  volume = {53},
  number = {3},
  pages = {185-216},
  doi = {10.1002/sapm1974533185},
  year = {1974},
}

@article{dCP76,
title = {A Characteristic Free Approach to Invariant Theory},
journal = {Advances in Mathematics},
volume = {21},
number = {3},
pages = {330-354},
year = {1976},
issn = {0001-8708},
doi = {10.1016/S0001-8708(76)80003-5},
author = {de Concini, Corrado and Procesi, Claudio}
}

@article{dCEP80,
  author = {de Concini, Corrado and Eisenbud, David and Procesi, Claudio},
  title = {Young Diagrams and Determinantal Varieties},
  journal = {Inventiones mathematicae},
  volume = {56},
  year = {1980},
  number = {2},
  pages = {129--165},
  doi = {10.1007/BF01392548},
}

@InProceedings{BC03,
  title={Gr{\"o}bner Bases and Determinantal Ideals},
  author={Bruns, Winfried and Conca, Aldo},
  booktitle="Commutative Algebra, Singularities and Computer Algebra",
  pages={9--66},
  year={2003},
  publisher={Springer Netherlands},
  doi={10.1007/978-94-007-1092-4_2},
}

@article{HT92,
  title = {Gröbner Bases and Multiplicity of Determinantal and Pfaffian Ideals},
  journal = {Advances in Mathematics},
  volume = {96},
  number = {1},
  pages = {1-37},
  year = {1992},
  doi = {10.1016/0001-8708(92)90050-U},
  author = {Jürgen Herzog and Ngô Viêt Trung},
}

@incollection{Mus03,
  title={The Development of Standard Monomial Theory-I},
  author={Musili, Chitikila},
  booktitle={A Tribute to C. S. Seshadri: Perspectives in Geometry and Representation Theory},
  pages={385--420},
  year={2003},
  publisher={Springer},
  doi={10.1007/978-93-86279-11-8_24},
}

@book{LR08,
  title={Standard Monomial Theory: Invariant Theoretic Approach},
  author={Lakshmibai, Venkatramani and Raghavan, Komaranapuram N.},
  year={2008},
  publisher={Springer},
  doi={10.1007/978-3-540-76757-2},
}

@book{Ses16,
  title={Introduction to the Theory of Standard Monomials},
  author={Seshadri, C. S.},
  year={2016},
  publisher={Springer},
  edition = {Second Edition},
  doi={10.1007/978-981-10-1813-8},
}

@book{Bur00,
  title={Completeness and Reduction in Algebraic Complexity Theory},
  author={B{\"u}rgisser, Peter},
  year={2000},
  publisher={Springer},
  doi={10.1007/978-3-662-04179-6},
}

@article{LST25,
  title={Superpolynomial Lower Bounds Against Low-Depth Algebraic Circuits},
  author={Limaye, Nutan and Srinivasan, Srikanth and Tavenas, S{\'e}bastien},
  journal={Journal of the ACM},
  volume={72},
  number={4},
  pages={1--35},
  year={2025},
  publisher={ACM},
  doi={10.1145/3734215},
}

@misc{BLRS25,
      title={New Bounds for the Ideal Proof System in Positive Characteristic}, 
      author={Amik Raj Behera and Nutan Limaye and Varun Ramanathan and Srikanth Srinivasan},
      year={2025},
      eprint={2506.16397},
      archivePrefix={arXiv}, 
}

@inproceedings{For24,
  title={Low-Depth Algebraic Circuit Lower Bounds over Any Field},
  author={Forbes, Michael A.},
  booktitle={39th Computational Complexity Conference (CCC 2024)},
  pages={31--1},
  year={2024},
  series =	{Leibniz International Proceedings in Informatics (LIPIcs)},
  publisher =	{Schloss Dagstuhl -- Leibniz-Zentrum f{\"u}r Informatik},
  doi={10.4230/LIPIcs.CCC.2024.31},
}

@inproceedings{Kal86,
  title={Uniform Closure Properties of P-Computable Functions},
  author={Kaltofen, Erich},
  booktitle={Proceedings of the 18th Annual ACM Symposium on Theory of Computing},
  pages={330--337},
  year={1986},
  doi={10.1145/12130.12163},
}

@inproceedings{Kal87,
  title={Single-Factor Hensel Lifting and Its Application to the Straight-Line Complexity of Certain Polynomials},
  author={Kaltofen, Erich},
  booktitle={Proceedings of the 19th Annual ACM Symposium on Theory of Computing},
  pages={443--452},
  year={1987},
  doi={10.1145/28395.28443},
}

@inproceedings{Kal89,
  author       = {Kaltofen, Erich},
  title        = {Factorization of Polynomials Given by Straight-Line Programs},
  booktitle      = {Randomness and Computation, Volume 5 of Advances in Computing Research},
  pages        = {375--412},
  year         = {1989},
}

@article{CKS19,
  title={Closure Results for Polynomial Factorization},
  author={Chou, Chi-Ning and Kumar, Mrinal and Solomon, Noam},
  journal={Theory of Computing},
  volume={15},
  number={1},
  pages={1--34},
  year={2019},
  doi={DOI: 10.4086/toc.2019.v015a013},
}

@misc{BKRRSS25,
      title={Closure Under Factorization From a Result of Furstenberg}, 
      author={Somnath Bhattacharjee and Mrinal Kumar and Shanthanu S. Rai and Varun Ramanathan and Ramprasad Saptharishi and Shubhangi Saraf},
      year={2025},
      eprint={2506.23214},
      archivePrefix={arXiv},
}

@article{ST21,
  title={Factorization of Polynomials Given by Arithmetic Branching Programs},
  author={Sinhababu, Amit and Thierauf, Thomas},
  journal={computational complexity},
  volume={30},
  number={2},
  pages={15},
  year={2021},
  publisher={Springer},
  doi={10.1007/s00037-021-00215-0},
}

@article{DSS22,
  title={Discovering the Roots: Uniform Closure Results for Algebraic Classes Under Factoring},
  author={Dutta, Pranjal and Saxena, Nitin and Sinhababu, Amit},
  journal={Journal of the ACM},
  volume={69},
  number={3},
  pages={1--39},
  year={2022},
  publisher={ACM},
  doi={10.1145/3510359},
}

@article{Oli16,
  title={Factors of Low Individual Degree Polynomials},
  author={Oliveira, Rafael},
  journal={computational complexity},
  volume={25},
  number={2},
  pages={507--561},
  year={2016},
  publisher={Springer},
  doi={10.1007/s00037-016-0130-2},
}

@article{Bur04,
  title={The Complexity of Factors of Multivariate Polynomials},
  author={B{\"u}rgisser, Peter},
  journal={Foundations of Computational Mathematics},
  volume={4},
  number={4},
  pages={369--396},
  year={2004},
  publisher={Springer},
  doi={10.1007/s10208-002-0059-5},
}

@article{NW94,
  title={Hardness vs Randomness},
  author={Nisan, Noam and Wigderson, Avi},
  journal={Journal of computer and System Sciences},
  volume={49},
  number={2},
  pages={149--167},
  year={1994},
  publisher={Elsevier},
  doi={10.1016/S0022-0000(05)80043-1},
}

@article{LL89,
  title={On the Order of Approximation in Approximative Triadic Decompositions of Tensors},
  author={Lehmkuhl, Thomas and Lickteig, Thomas},
  journal={Theoretical computer science},
  volume={66},
  number={1},
  pages={1--14},
  year={1989},
  publisher={Elsevier},
  doi={10.1016/0304-3975(89)90141-2},
}

@InProceedings{DGIJL24,
  author =	{Dutta, Pranjal and Gesmundo, Fulvio and Ikenmeyer, Christian and Jindal, Gorav and Lysikov, Vladimir},
  title =	{{Fixed-Parameter Debordering of Waring Rank}},
  booktitle =	{41st International Symposium on Theoretical Aspects of Computer Science (STACS 2024)},
  pages =	{30:1--30:15},
  series =	{Leibniz International Proceedings in Informatics (LIPIcs)},
  year =	{2024},
  publisher =	{Schloss Dagstuhl -- Leibniz-Zentrum f{\"u}r Informatik},
  doi = {10.4230/LIPIcs.STACS.2024.30},
}

@inproceedings{DDS22,
  title={Demystifying the Border of Depth-3 Algebraic Circuits},
  author={Dutta, Pranjal and Dwivedi, Prateek and Saxena, Nitin},
  booktitle={2021 IEEE 62nd Annual Symposium on Foundations of Computer Science (FOCS)},
  pages={92--103},
  year={2022},
  organization={IEEE},
  doi={10.1109/FOCS52979.2021.00018},
}

@inproceedings{BDS24,
  title={Learning the Coefficients: A Presentable Version of Border Complexity and Applications to Circuit Factoring},
  author={Bhargav, C. S. and Dwivedi, Prateek and Saxena, Nitin},
  booktitle={Proceedings of the 56th Annual ACM Symposium on Theory of Computing},
  pages={130--140},
  year={2024},
  doi={10.1145/3618260.3649743},
}

@misc{Shp25,
      title={Improved Debordering of Waring Rank}, 
      author={Amir Shpilka},
      year={2025},
      eprint={2502.03150},
      archivePrefix={arXiv},
}

@book{DEP82,
     author = {de Concini, Corrado and Eisenbud, David and Procesi, Claudio},
     title = {Hodge Algebras},
     series = {Ast\'erisque},
     publisher = {Soci\'et\'e math\'ematique de France},
     number = {91},
     year = {1982},
     mrnumber = {680936},
     zbl = {0509.13026},
     url = {https://www.numdam.org/item/AST_1982__91__1_0/},
}

@article{VV86,
  title={NP is as Easy as Detecting Unique Solutions},
  author={Valiant, Leslie G. and Vazirani, Vijay V.},
  journal={Theoretical Computer Science},
  volume={47},
  number={1},
  pages={85--93},
  year={1986},
  publisher={Elsevier},
  doi={10.1016/0304-3975(86)90135-0},
}

@article{MVV87,
  title={Matching is as Easy as Matrix Inversion},
  author={Mulmuley, Ketan and Vazirani, Umesh V. and Vazirani, Vijay V.},
  journal={Combinatorica},
  volume={7},
  number={1},
  pages={105--113},
  year={1987},
  publisher={Springer},
  doi={10.1007/BF02579206},
}

@article{FGT19,
  title={Bipartite Perfect Matching Is in Quasi-NC},
  author={Fenner, Stephen and Gurjar, Rohit and Thierauf, Thomas},
  journal={SIAM Journal on Computing},
  volume={50},
  number={3},
  pages={STOC16-218--STOC16-235},
  year={2019},
  publisher={SIAM},
  doi={10.1137/16M1097870},
}

@inproceedings{ST17,
  title={The Matching Problem in General Graphs Is in Quasi-NC},
  author={Svensson, Ola and Tarnawski, Jakub},
  booktitle={2017 IEEE 58th Annual Symposium on Foundations of Computer Science (FOCS)},
  pages={696--707},
  year={2017},
  organization={IEEE},
  doi={10.1109/FOCS.2017.70},
}

@article{AMS10,
  title={New Results on Noncommutative and Commutative Polynomial Identity Testing},
  author={Arvind, Vikraman and Mukhopadhyay, Partha and Srinivasan, Srikanth},
  journal={computational complexity},
  volume={19},
  number={4},
  pages={521--558},
  year={2010},
  publisher={Springer},
  doi={10.1007/s00037-010-0299-8},
}

@article{GT20,
  title={Linear Matroid Intersection Is in Quasi-NC},
  author={Gurjar, Rohit and Thierauf, Thomas},
  journal={computational complexity},
  volume={29},
  number={2},
  pages={9},
  year={2020},
  publisher={Springer},
  doi={10.1007/s00037-020-00200-z},
}

@article{GGR24,
  title={A Deterministic Parallel Reduction from Weighted Matroid Intersection Search to Decision},
  author={Ghosh, Sumanta and Gurjar, Rohit and Raj, Roshan},
  journal={Algorithmica},
  volume={86},
  number={4},
  pages={1057--1079},
  year={2024},
  publisher={Springer},
  doi={10.1137/1.9781611977073.44},
}

@article{BCGL92,
  title={On the Theory of Average Case Complexity},
  author={Ben-David, Shai and Chor, Benny and Goldreich, Oded and Luby, Michael},
  journal={Journal of Computer and System Sciences},
  volume={44},
  number={2},
  pages={193--219},
  year={1992},
  doi={10.1016/0022-0000(92)90019-F},
}

@book{BV88,
  title = {Determinantal Rings},
  publisher = {Springer},
  author = {Bruns, Winfried and Vetter, Udo},
  year = {1988},
  doi = {10.1007/BFb0080378},
}

@InProceedings{And20,
  author =	{Andrews, Robert},
  title =	{{Algebraic Hardness versus Randomness in Low Characteristic}},
  booktitle =	{35th Computational Complexity Conference (CCC 2020)},
  pages =	{37:1--37:32},
  series =	{Leibniz International Proceedings in Informatics (LIPIcs)},
  year =	{2020},
  publisher =	{Schloss Dagstuhl -- Leibniz-Zentrum f{\"u}r Informatik},
  doi =		{10.4230/LIPIcs.CCC.2020.37},
}

@article{HvMM24,
  doi = {http://dx.doi.org/10.4086/toc.2024.v020a001},
  author = {Hu, Ivan and van Melkebeek, Dieter and Morgan,  Andrew},
  title = {Polynomial Identity Testing via Evaluation of Rational Functions},
  journal = {Theory of Computing},
  year = {2024},
  volume={20},
  number={1},
}

@article{BSV20,
  author       = {Vishwas Bhargava and
                  Shubhangi Saraf and
                  Ilya Volkovich},
  title        = {Deterministic Factorization of Sparse Polynomials with Bounded Individual
                  Degree},
  journal      = {Journal of the {ACM}},
  volume       = {67},
  number       = {2},
  pages        = {8:1--8:28},
  year         = {2020},
  doi          = {10.1145/3365667},
}

\end{document}